\documentclass[11pt]{article}

\usepackage{bm, commath}
\usepackage{natbib}
\usepackage{caption}
\usepackage{graphicx}
\usepackage{subcaption}
\usepackage{amsmath, amsfonts, amsthm}
\usepackage{float}
\usepackage{booktabs,siunitx}
\usepackage{url}
\usepackage{multirow}
\usepackage{amssymb}
\usepackage{bm}
\usepackage{mathrsfs}
\usepackage{xr}
\usepackage{authblk}

\usepackage{listings}
\allowdisplaybreaks
\usepackage{bbm}
\usepackage{textcomp}
\usepackage[margin=1in]{geometry}
\usepackage{authblk}
\usepackage[ruled]{algorithm2e}
\SetKwInput{KwParam}{Parameter}
\SetAlgoCaptionLayout{centerline}

\usepackage{sectsty}
\usepackage[onehalfspacing]{setspace}
\usepackage[usenames,dvipsnames]{color}
\usepackage[utf8]{inputenc}
\usepackage{float}
\usepackage{tikz}
\usetikzlibrary{trees}
\usetikzlibrary{shapes,decorations,arrows,calc,arrows.meta,fit,positioning}
\usetikzlibrary{shapes.multipart}
\tikzset{
    -Latex,auto,node distance =1 cm and 1 cm,semithick,
    state/.style ={ellipse, draw, minimum width = 0.7 cm},
    state1/.style ={ draw, minimum width = 0.7 cm},
    point/.style = {circle, draw, inner sep=0.04cm,fill,node contents={}},
    bidirected/.style={Latex-Latex,dashed},
    el/.style = {inner sep=2pt, align=left, sloped}
}

\newtheorem{theorem}{Theorem}

\newtheorem{lemma}{Lemma}

\usepackage{mathtools}

\usepackage{xcolor}

\makeatletter
\renewcommand{\algocf@captiontext}[2]{#1\algocf@typo. \AlCapFnt{}#2} 
\def\@algocf@capt@plain{top}
\renewcommand{\algocf@makecaption}[2]{%
  \addtolength{\hsize}{\algomargin}%
  \sbox\@tempboxa{\algocf@captiontext{#1}{#2}}%
  \ifdim\wd\@tempboxa >\hsize
  \hskip .5\algomargin%
  \parbox[t]{\hsize}{\algocf@captiontext{#1}{#2}}
  \else%
  \global\@minipagefalse%
  \hbox to\hsize{\box\@tempboxa}
  \fi%
  \addtolength{\hsize}{-\algomargin}%
}
\makeatother

\def\hat{\widehat}
\def\tilde{\widetilde} 
\def\P{{\rm pr}}

\def\var{{\rm var}}
\def\cov{{\rm cov}}
\def\olim{{O}}

\def\ohat{\hat{O}}

\begin{document}

\sectionfont{\bfseries\large\sffamily}%

\subsectionfont{\bfseries\sffamily\normalsize}%

\title{
	Covariate-Adjusted Log-Rank Test:  Guaranteed Efficiency Gain and Universal Applicability}

\author{Ting Ye}
\affil{Department of Biostatistics, University of Washington,  Seattle, Washington 98195, U.S.A.\thanks{tingye1@uw.edu}}

\author{Jun Shao}
\affil{School of Statistics, East China Normal University, Shanghai 200241, China\\ Department of Statistics, University of Wisconsin, Madison, Wisconsin 53706, U.S.A.\thanks{shao@stat.wisc.edu}}

\author{Yanyao Yi}
\affil{Global Statistical Sciences, Eli Lilly and Company, 
	Indianapolis, Indiana  46285, U.S.A.\thanks{yi\_yanyao@lilly.com}}

\maketitle

\begin{abstract}
	Nonparametric covariate adjustment is considered for log-rank type tests of treatment effect with  right-censored time-to-event data from clinical trials applying covariate-adaptive randomization. 
	Our proposed covariate-adjusted log-rank test has a simple explicit formula and  
	a guaranteed efficiency gain over the unadjusted  test.
	We also show that our proposed  test achieves universal applicability in the sense that the same formula of 
	test can be universally applied to simple randomization and all commonly used covariate-adaptive randomization schemes such as the stratified permuted block and  Pocock and Simon's minimization, 
	which is not a property enjoyed by the unadjusted log-rank test. 
	Our method is supported by novel asymptotic theory and empirical results 
	for type I error and power of tests. 
\end{abstract}


\clearpage

\section{Introduction} 
\label{sec: intro}

In clinical trials, adjusting for baseline covariates has been widely advocated as a way to improve efficiency for demonstrating 
treatment effects,
``under approximately the same minimal statistical assumptions that would be needed for unadjusted estimation'' 
\citep{ICHE9,ema:2015aa,fda:2019aa}. In testing for effect between two treatments with right-censored time-to-event outcomes, 
adjusting for covariates using the Cox  proportional hazards model has been demonstrated to yield valid tests even if the Cox model is misspecified \citep{Lin:1989aa, KONG:1997aa, DiRienzo:2002aa}. However,  these tests  may be less powerful than the log-rank test that does not adjust for any covariates 
when the Cox model  is misspecified \citep{KONG:1997aa}. 
Although efforts have been made to improve the efficiency of the log-rank test through covariate adjustment from semiparametric theory \citep{Lu:2008,moore2009increasing}, the solutions are  complicated  and their validities are established only under simple randomization, i.e., treatments are assigned to patients completely at random. 

To balance the number of patients in each treatment arm across baseline prognostic factors in clinical trials with  sequentially arrived patients, 
covariate-adaptive randomization has become the new norm.  
From 1989 to 2008, covariate-adaptive randomization was used in more than 500 clinical trials 
\citep{Taves2010};  
among nearly 300  trials published in two years, 2009 and 2014, 
237 of them applied covariate-adaptive randomization \citep{Ciolino:2019aa}.  The two most popular covariate-adaptive randomization schemes are the stratified permuted block \citep{Zelen:1974aa} 
and Pocock-Simon's minimization \citep{Taves:1974aa, Pocock:1975aa}. Other schemes can be found in two reviews, \cite{Schulz:2018aa} and \cite{Shao:2021aa}. 
Unlike simple randomization, covariate-adaptive randomization generates a dependent sequence of treatment assignments,     which may render conventional methods developed under simple randomization not necessarily valid under covariate-adaptive randomization
\citep{ema:2015aa,fda:2019aa}. 
For time-to-event data under covariate-adaptive randomization,	\cite{Ye:2020survival} 
shows that  some conventional tests including the log-rank test
are  conservative and \cite{Wang:2021wg} shows that the Kaplan-Meier estimator of survival function has reduced variance compared to that under simple randomization.


The discussion so far has brought up two issues 
in adjusting covariates, the guaranteed efficiency gain over unadjusted method under the same assumption and the wide applicability  to all commonly used covariate-adaptive randomization schemes. 
These  issues have been well addressed
when adjustments are made under  linear working models  for  non-time-to-event data
\citep{Tsiatis:2008aa,zhang2008improving, Lin:2013aa, ye2021better}. \cite{ye2021better} also shows that adjustment via linear working   models can achieve universal applicability in the sense that the same inference procedure can be universally applied to all commonly used covariate-adaptive randomization schemes, a desirable property for application. For right-censored time-to-event outcomes,
to the best of our knowledge, no result has been established for covariate adjustment with guaranteed efficiency gain and universal applicability.



In this paper we propose a nonparametric  covariate adjustment  method for the  log-rank test, which has a simple explicit form and can achieve the goal of guaranteed efficiency gain over the unadjusted  log-rank test 	as well as 	universal applicability 
to simple randomization and all commonly used covariate-adaptive randomization schemes. 
Note that the unadjusted log-rank test is not valid under  covariate-adaptive randomization; although it can be modified to be applicable to some randomization schemes 
\citep{Ye:2020survival}, the modification needs to be tailored to each randomization scheme, i.e., 
no	universal applicability.
Our  main idea is to  obtain a particular ``derived outcome'' for each patient  from linearizing the log-rank test statistic and then apply the generalized regression adjustment  or  augmentation \citep{Cassel:1976aa,Lu:2008, Tsiatis:2008aa,zhang2008improving} to the derived outcomes. We also develop parallel results for the stratified log-rank test with adjustment for additional covariates. Our proposed tests are supported by novel asymptotic theory of the existing and proposed statistics under null hypothesis and alternative without requiring any specific model assumption, and under all commonly used covariate-adaptive randomization schemes.
Estimation and confidence intervals for treatment effects 
after testing are also discussed.
Our theoretical results are corroborated by a simulation study that examines finite sample type I error and power of tests.  
A real data example is included for illustration. 

\vspace{-1mm}

\section{Preliminaries}

For a patient from the population under investigation, 
let $T_{j}$ and  $C_{j}$ be the \emph{potential} failure time and right-censoring time, respectively, under treatment $j=0 $ or 1, and $W$ be a vector containing all  baseline covariates and time-varying covariates, observed or unobserved. 
Suppose that a random sample of $n$ patients is obtained from the population with independent 	$({T}_{i0} ,C_{i0}, {T}_{i1}, C_{i1},W_i)$, $i=1,... , n$,  identically distributed as $(T_0, C_0, T_1, C_1, W)$. 
For each patient, only one of the two treatments is received. Thus, if patient $i$ receives treatment $j$, then the observed outcome with possible right censoring is $ \{ \min ({T}_{ij}, C_{ij} ) , \delta_{ij} \}$,  where 	$\delta_{ij}$ is the indicator of the event ${T}_{ij} \leq C_{ij}$.

Let $I_i$ be a binary treatment indicator for patient $i$ and 
$0< \pi <1 $ be the pre-specified treatment assignment proportion for treatment 1. 
Consider the design, i.e., the generation of $I_i$'s for $n$ sequentially arrived patients.  Simple randomization assigns patients to treatments completely at random with $\P(I_i =1 ) = \pi$ for all $i$, which does not make use of baseline covariates and  may yield  treatment proportions that substantially deviate from the target $\pi$ across levels of some
prognostic factors. 
Because of this, covariate-adaptive randomization using a sub-vector  $Z$ of the baseline covariates in $W$ is widely applied, which 	does not use any model and  is  nonparametric. 
All  commonly used covariate-adaptive randomization schemes
satisfy the following mild condition \citep{Baldi-Antognini:2015aa}.
\begin{description} 
	\item (D) 	The covariate $Z$ 	for which we want to balance in treatment assignment  is an observed discrete baseline covariate with finitely many joint levels;	 conditioned on $(Z_i,\,  i\!=\!1,..., n)$, $(I_i,\,  i\!=\!1,..., n)$ is  conditionally independent of 
	$({T}_{i1}, C_{i1}, {T}_{i0}, C_{i0},  W_i, \, i\!=\!1,..., n)$; $E(I_i \mid Z_1,..., Z_n) = \pi$ for all $i$; and 
	for every level $z$ of $Z$, $n_{z1} /n_z \rightarrow  \pi $ in probability as $n\to \infty$, where $n_z$
	is the number of patients with $Z_i=z$ and $n_{z1}$ is the number of
	patients with $Z_i=z$ and $I_i=1$. 
\end{description}
Although simple randomization is not counted as covariate-adaptive randomization, it also satisfies (D).  

We focus on testing the following null hypothesis of no treatment effect, which is the null hypothesis 
when the conventional log-rank test is applied, $H_0: \lambda_1(t)=\lambda_0(t)$ for any time $t$, versus the alternative that $H_0$ does not hold, where $\lambda_j(t)$ is the  unspecified hazard function of $T_j$, unconditional on covariates. 

After data are collected from all patients, 		a test statistic ${\cal T}$  is a function of  observed data,  constructed such that  $H_0$ is rejected if and only if $|{\cal T}|>z_{\alpha/2}$, where $\alpha$ is a given significance level and $z_{\alpha/2}$ is the $(1-\alpha/2)$th quantile of the standard normal distribution. A test ${\cal T}$ is (asymptotically) valid if under $H_0$, $	\lim_{n\rightarrow \infty} \P (|{\cal T}|>z_{\alpha/2})\leq \alpha$,
with equality holding for at least one parameter value under the null hypothesis $H_0$. A test ${\cal T}$ is (asymptotically) conservative if under $H_0$, there exists an $\alpha_0$ such that $	\lim_{n\rightarrow \infty} \P (|{\cal T}|>z_{\alpha/2})\leq \alpha_0 < \alpha$.

The test statistic of  log-rank test is \vspace{-2mm}
\begin{equation}
	{\cal T}_{\mathrm L}  = \sqrt{n} \, \hat U_{\mathrm L} / \, \hat\sigma_{\mathrm L}   \label{logrank} 
\end{equation}
\citep{mantel1966evaluation,kalbfleisch2011statistical},
where 
$$
\hat	U_{\mathrm L}   =  
\frac{1}{{n}} \sum_{i=1}^{n}\int_0^\tau \left\{I_i-\frac{\bar{Y}_{1}(t)}{\bar Y(t)}\right\}dN_i(t), \qquad 
\hat{\sigma}^2_{\mathrm L}   =\frac{1}{n}\sum_{i=1}^n\int_0^\tau \frac{\bar Y_1(t)\bar Y_0(t)}{\bar Y(t)^2}dN_i(t) ,
$$
$\bar Y_1 (t)=\sum_{i=1}^n I_i Y_i(t)/n, \bar Y_0 (t)=\sum_{i=1}^n(1-I_i)Y_i(t)/n $, $\bar Y(t) = \bar Y_1(t)+\bar  Y_0 (t)$,  $Y_i(t)=I_iY_{i1}(t) \, + $ $(1-I_i)Y_{i0}(t)$, 
$Y_{ij}(t)= $ the indicator of the event $\min ({T}_{ij}, C_{ij})\geq  t$,  $ N_i(t) = I_i N_{i1} (t) $ $+ (1-I_i)  N_{i0} (t)$ is the counting process of observed failures, 
$ N_{ij}(t) $ is the indicator of the event ${T}_{ij} \leq \min (t, C_{ij} ) $,  and the upper limit $\tau$ in the integral is a point satisfying $\P\big(\min ({T}_{ij}, C_{ij}) \geq \tau \big)>0$ for $j=0,1$.

The log-rank test ${\cal T}_{\mathrm L}$ in (\ref{logrank})  is valid under simple randomization
and the following assumption: 
\begin{description}
	\item (C) $ \ C_I \perp (T_I,W) \mid I$, where $I$ is the treatment indicator, 
	$\perp$ denotes independence, and $\mid$ is conditioning. 
\end{description} 
Under (C), censoring can be affected by treatment, but not by any covariate and, hence, it is termed as non-informative censoring. It is a strong assumption on censoring, but is required in order to have a valid  
nonparametric  log-rank test without requiring any model on $T_j$ or $C_j$ \citep{KONG:1997aa,DiRienzo:2002aa, Lu:2008, parast2014landmark, zhang2015robust}.

As ${\cal T}_{\mathrm L}$  does not utilize any baseline covariate information,  it is used as 
the benchmark in 
considering baseline covariate adjustment for efficiency gain, 
under the same  assumption (C) ``that would be needed for unadjusted" ${\cal T}_{\mathrm L}$ \citep{fda:2019aa}.

For the validity of log-rank test and our covariate-adjusted log-rank test proposed in \S3, condition (C) can be weakened to  the following (CR), but a mild  transformation model assumption (TR)  is needed.
\begin{description}
	\item
	(CR) There  is  a  possibly time-varying covariate vector $V\subset W$ such that 
	$C_I \perp (T_I ,W) \mid (I , V )$ and the ratio $\P(C_1 \geq t \mid V)  / \P(C_0 \geq t\mid V)$ is a function of $t$ only. \vspace{2mm}
	\item  (TR)  There exists an increasing function $h$ such that 
	$h(\P (T_0 \geq t \mid V)) = \theta + h(\P (T_1 \geq t \mid V))$ for all $(t,V)$ and a constant $\theta$, where $V$ is in (CR) and both $h$ and $\theta$ are unknown. 
\end{description} 
The first condition  in (CR), $C_I \perp (T_I ,W) \mid (I , V )$, is
weaker than what is typically assumed as it only requires the existence of  $V$ that  can be not fully observed.
The  condition in (CR) about the ratio $\P(C_1 \geq t \mid V)  / \P(C_0 \geq t\mid V)$ 
is interpreted by  \cite{KONG:1997aa} as no treatment-by-$V$ interaction on censoring. It is plausible if censoring is purely administrative  at a fixed calendar time  while patients enter the study randomly depending on $V$, or censoring is due to side-effects  related to the treatment  but not  $V$.
The transformation model (TR) is a general model discussed in \cite{cheng1995analysis}, which  includes many commonly used semiparametric models as special cases, e.g.,  the Cox proportional hazards model with $h (s) = - \log (- \log (s))$.

There is also a line of research weakening (C) to  censoring-at-random \citep{robins2000correcting, lu2011semiparametric, diaz2019improved}  under which, however,  the log-rank test is not valid and needs to be replaced by a weighted 
log-rank test that requires a correctly specified censoring distribution as the weights 
are inverse probabilities of censoring.
Thus, the conditions and properties of weighted log-rank tests are not comparable with those of the log-rank test. 
Furthermore,  the validity of weighted  log-rank tests has only been established under simple randomization. The study of weighted log-rank tests
under covariate-adaptive randomization  is a future work. 

\section{Covariate-Adjusted Log-Rank Test}

Let  $X \subset W$ contains 
observed baseline covariates  to be adjusted in the construction of tests, 
with a nonsingular covariance matrix 
$ \Sigma_X = \var (X)$. 
In this section, we develop a nonparametric covariate-adjusted log-rank test that has a simple and explicit formula,	enjoys guaranteed efficiency gain over the log-rank test, and is universally valid under all covariate-adaptive randomization schemes  satisfying (D). 

To develop our covariate adjustment method, we first consider the following linearization of  $\hat U_{\mathrm L}$  in (\ref{logrank}),  \vspace{-1mm}
\begin{align*}
	\hat U_{\mathrm L} & =U_{\mathrm {lin}}  + n^{-1/2}\, o_p(1) , \quad \quad \ \ 
	U_{\mathrm {lin}}  = \frac{1}{n} \sum_{i=1}^{n} \{  I_i   \olim_{i1} -  (1-I_i) \olim_{i0} \}  ,   \vspace{-4mm} \\
	\olim_{ij} & =\int_0^\tau  \{ 1- \mu(t)\}^j \{\mu(t)\}^{1-j} 
	\{dN_{ij}(t)-Y_{ij}(t)  p(t) dt\}, \quad j=0,1,  
\end{align*} 
\citep{Lin:1989aa, Ye:2020survival}, 
where $\mu(t)  = E  (I_i  \mid Y_i(t)= 1)  $, 	$p(t) dt = {E}\{d N_i(t)\}/E\{Y_i(t)\}$, and 
$o_p(1)  $ denotes a term converging to 0 in probability as  $n \to \infty $. Note that 
$U_{\mathrm {lin}} $ is an average of random variables that are independent and identically distributed  under simple  randomization. If we treat $\olim_{ij}$'s in $U_{\mathrm {lin}}$ as ``outcomes" and  apply the generalized regression adjustment   or augmentation \citep{Cassel:1976aa, Tsiatis:2008aa}, then we obtain the following  covariate-adjusted ``statistic", \vspace{-1mm}
\begin{align}
	\begin{split}
		U_{\mathrm {Clin}} 
		&=  \frac{1}{n} \sum_{i=1}^{n} \left[ I_i  \{ \olim_{i1} - (X_i- \bar X)^\top \beta_1 \} -  (1-I_i) \{ \olim_{i0}  - (X_i- \bar X)^\top \beta_0\} \right]  \vspace{-4mm} \\
		& =U_{\mathrm {lin}}  	- \frac{1}{n}\sum_{i=1}^{n}
		\left\{ I_i (X_i -\bar X)^\top \beta_1 - (1-I_i) (X_i - \bar X)^\top \beta_0 \right\} ,  \label{CL}
	\end{split}
\end{align} 
where 	$ \bar X $ is the sample mean of all $ X_i $'s, $a^\top$ is the transpose of vector $a$, and 
$\beta_j = \Sigma_X^{-1} \cov ( X_i, \olim_{ij} )$ for $j=0,1$.
Because the distribution of baseline covariate $X_i$ is not affected by treatment,  the last term on the right hand side of \eqref{CL} has mean 0. Under simple randomization, it follows from the theory of generalized regression \citep{Cassel:1976aa} that
$\var (U_{\mathrm {Clin}} ) \leq \var ( U_{\mathrm {lin}} )$, and thus the  covariate-adjusted $U_{\mathrm {Clin}} $ in \eqref{CL} has a guaranteed efficiency gain over the unadjusted $U_{\mathrm {lin}} $. This also holds under covariate-adaptive randomization; see Theorem S1 of  Supplementary Material. 

To derive our covariate-adjusted procedure, it remains to find appropriate statistics 
to replace $\olim_{ij}$'s and $\beta_j$'s in (\ref{CL}) because they involve unknown quantities. We consider the following sample analog of $\olim_{ij}$, \vspace{-1mm}
\begin{align}
	\ohat_{ij}  = \int_{0}^\tau \frac{ \bar Y_{1-j} (t)}{ \bar Y(t)} \left\{  dN_{ij} (t) -    Y_{ij} (t)  \frac{d\bar  N(t) }{ \bar Y(t) } \right\}  , \quad j=0,1,  \label{hatoij}
\end{align}
where 	$\bar N(t)= \sum_{i=1}^{n} N_i(t)/n$. Using  a correct form of $\hat O_{ij}$ is 
important, as it captures the true correlation between $\olim_{ij} $ and $X_i$. See the discussion after Theorem S1 of  Supplementary Material.
Replacing  $\olim_{ij}$ in \eqref{CL} by the derived outcome $\ohat_{ij}  $ in (\ref{hatoij}), we obtain the following covariate-adjusted version of $\hat U_{\mathrm L}$, 
\begin{align}
\begin{split}
\hat  U_{\mathrm {CL}} 
&=  \frac{1}{n}\sum_{i=1}^{n} \left[ I_i \{  \ohat_{i1} -  (X_i- \bar X)^\top \hat\beta_1  \} - (1-I_i)  \{   \ohat_{i0}  - ( X_i- \bar X)^\top \hat \beta_0  \}  \right] \vspace{-4mm}
\\
& =\hat U_L
- \frac{1}{n}\sum_{i=1}^{n}
\left\{ I_i (X_i -\bar X)^\top \hat\beta_1 - (1-I_i) (X_i - \bar X)^\top \hat\beta_0 \right\} ,  
\end{split} \label{CL1}
\end{align}
where  the last equality  follows from the algebraic  identity $ \hat	U_{\mathrm L}  = n^{-1} \sum_{i=1}^{n} \{ I_i \ohat_{i1}- (1-I_i) \ohat_{i0}\}$, 
\begin{equation}\label{hatbeta}
\hat\beta_j =    \bigg\{  \sum_{i: I_i= j} (X_i -  \bar X_{j}  ) (X_i -  \bar X_{j}  )^\top   \bigg\}^{-1}    \sum_{i: I_i= j} (X_i -  \bar X_{j}  )  \ohat_{ij} 
\end{equation}
is a sample analog of $\beta_j = \Sigma_X^{-1} \cov ( X_i, \olim_{ij} )$, and
$\bar X_{j} $ is the sample mean of $X_i$'s  with $I_i=j$. 
By Lemma S1
in  Supplementary Material, 
$\hat \beta_j$ in (\ref{hatbeta}) converges to $\beta_j $ in probability, which 
guarantees that $\hat U_{\mathrm {CL}}$ in (\ref{CL1}) reduces the variability of $\hat U_{\mathrm L}$ in (\ref{logrank}). 
Thus,	we propose the following covariate-adjusted log-rank test,  
\begin{align}
{\cal T}_{\mathrm {CL}}  = \sqrt{n} \ \hat U_{\mathrm {CL}} /\,  \hat\sigma_{\mathrm {CL}},  \label{eq: TCL} 
\end{align}
where  
$ \hat\sigma_{\mathrm {CL}}^2 = \hat\sigma_{\mathrm L}^2 - \pi(1-\pi) ( \hat\beta_1 + \hat\beta_0 )^\top  \hat \Sigma_X  ( \hat\beta_1 + \hat\beta_0 ) $ whose form is suggested by $\sigma^2_{\mathrm {CL}}$  in Theorem 1, $ \hat\sigma_{\mathrm L}^2 $ is defined in (\ref{logrank}), and $\hat \Sigma_X $ is the sample covariance matrix of all $X_i$'s.

Asymptotic properties of covariate-adjusted log-rank test ${\cal T}_{\mathrm {CL}}$ in (\ref{eq: TCL}) are established in the following theorem. All technical proofs are in  Supplementary Material. 
In what follows, $\xrightarrow{d}  $ or $\xrightarrow{p}  $ 
denotes convergence in distribution or probability, as $n \to \infty$.

\begin{theorem}\label{theo: 1}
Assume {(C)} or {(CR)-(TR)}. Assume also {(D)} and that all levels of $ Z_i $  used in covariate-adaptive randomization are included in $ X_i $ as a sub-vector. Then, the following results hold regardless of which covariate-adaptive randomization scheme is applied. \vspace{-3mm}
\begin{itemize}
\item[(a)]  Under the null $H_0$ or  alternative hypothesis, 
$
\sqrt{n}  \{ \hat  U_{\mathrm {CL}} - ( n_1 \theta_1- n_0\theta_0)/n  \}  \xrightarrow{d}  N\left(0,\, \sigma_{\mathrm {CL}}^2 \right)$, 
where  $\theta_j = E(\olim_{ij})$, 
$n_j =$ the number of patients in treatment $j$,   
$	\sigma_{\mathrm {CL}}^2   =  \sigma^2_{\mathrm L} - \pi(1-\pi) (\beta_1+ \beta_0)^\top \Sigma_X (\beta_1+\beta_0) $, and 
$		\sigma^2_{\mathrm L}  = \pi \var (\olim_{i1} ) + (1-\pi) \var (\olim_{i0} ) $.
\item[(b)] Under the null hypothesis $H_0$, \ $\theta_1=\theta_0 = 0$, 
$\ \hat\sigma^2_{\mathrm {CL}} \xrightarrow{p} \sigma^2_{\mathrm {CL}}$, and $  
\	{\cal T}_{\mathrm {CL}} \xrightarrow{d} N(0,1)$, 
i.e., 	${\cal T}_{\mathrm {CL}}$ is valid. 
\item[(c)] Under the local alternative hypothesis that  $\theta_j = c_j n^{-1/2}$ with 
$c_j$'s not depending on $n$ and that
$\lambda_1(t) / \lambda_0(t) $ is bounded and $\to 1$  for every $t$, 
$ {\cal T}_{\mathrm {CL}} \xrightarrow{d}  N( \{\pi c_1-(1-\pi)c_0\}/ \sigma_{\mathrm {CL}}, \ 1)$. 
\end{itemize}
\end{theorem}

The results under alternative hypothesis in Theorem 1 are obtained without any specific model on the distribution of $T_j$ or $C_j$, different from many published research  articles assuming a specific model under alternative such as the Cox proportional hazards model for $T_j$. 

Theorem 1  shows that 
${\cal T}_{\mathrm {CL}}$ in (\ref{eq: TCL}) is applicable to all  randomization schemes satisfying (D) with a universal formula, if all levels of $Z_i$ are included in  $X_i$. Tests with universal applicability are desirable for application, as the complication of using tailored formulas for different randomization schemes is avoided.

To show that ${\cal T}_{\mathrm {CL}}$ in (\ref{eq: TCL}) has a guaranteed efficiency gain over the benchmark ${\cal T}_{\mathrm L}$  in (\ref{logrank}), we  establish an asymptotic result for ${\cal T}_{\mathrm L}$ under covariate-adaptive randomization satisfying an additional condition (D$^\dag$): 
\begin{description}
\item
{(D$^\dag$) As $n\rightarrow\infty$, 
$\sqrt{n} \left( n_{z1}/ n_z - \pi, \, z\in \mathcal{Z} \right)^\top \mid Z_1,\dots, Z_n \xrightarrow{d} N\left(0,  
\Omega 
\right)$,   where $\mathcal{Z} $ is the set containing all levels of $Z$,  
$\Omega $ is the diagonal matrix 
whose diagonal entries are $  \nu / \P (Z=z)$, $z\in \mathcal{Z}$, and $\nu\leq \pi(1-\pi)$ is a known constant depending on the randomization scheme}. 
\end{description} 

\begin{theorem}  \label{theo: log-rank}
Assume {(C)} or {(CR)-(TR)}.  Assume also {(D)} and {(D$^\dag$)}. 
Then the following results hold. \vspace{-3mm}
\begin{itemize}
\item[(a)] Under the null $H_0$ or  alternative hypothesis, 
$\sqrt{n}  \{  \hat  U_{\mathrm L} -(n_1 \theta_1- n_0\theta_0)/n \}  \xrightarrow{d}  N\left(0,\, \sigma_{\mathrm L}^2 (\nu ) \right)
$, 
where $n_j$ and $\theta_j$ are given in Theorem 1,	$	\sigma_{\mathrm L}^2  (\nu ) = \sigma^2_{\mathrm L} 
- \{  \pi (1-\pi)- \nu \} \var  \{  E(\olim_{i1}| Z_i) + E(\olim_{i0}| Z_i)  \}  $
for $\nu$ given in {(D$^\dag$)}, and  $\sigma^2_{\mathrm L}  $ is defined in Theorem 1. 
\item[(b)] Under the null hypothesis $H_0$,   \ $\theta_1 = \theta_0 =0$, \ $\hat\sigma_{\mathrm L}^2  \xrightarrow{p} \sigma_{\mathrm L}^2 $, \ and \
${\cal T}_{\mathrm L} \xrightarrow{d} N\left(0,\sigma_{\mathrm L}^2(\nu ) / \sigma_{\mathrm L}^2\right)$. Hence,	${\cal T}_{\mathrm L}$ is conservative unless $\nu= \pi(1-\pi)$ or
$ E(O_{i1} | Z_i) + 
E(O_{i0} | Z_i)=0$  almost surely under $H_0$.
\item [(c)] Under the local alternative hypothesis  in Theorem 1(c), 
$ {\cal T}_{\mathrm L} \xrightarrow{d}  
N(  \{\pi c_1- (1-\pi )c_0\}/\sigma_{\mathrm L} , \ \sigma_{\mathrm L}^2(\nu ) /\sigma_{\mathrm L}^2  )$.
\end{itemize}
\end{theorem}

Under simple randomization, (D$^\dag$) holds with $\nu = \pi (1-\pi)$ and, hence,  
Theorem 2 also applies  with  
$\sigma^2_{\mathrm L}(\nu) = \sigma^2_{\mathrm L}$. Under the local alternative  specified in Theorem 1(c) with  $\pi c_1 - (1-\pi )c_0 \neq 0$,
by Theorems \ref{theo: 1}(c) and \ref{theo: log-rank}(c), 
Pitman's asymptotic relative efficiency of ${\cal T}_{\mathrm {CL}}$ in (\ref{eq: TCL}) with respect to the benchmark ${\cal T}_{\mathrm L}$ in (\ref{logrank}) is 
$  \sigma_{\mathrm L}^2/\sigma_{\mathrm {CL}}^2 = 1 + \pi(1-\pi) (\beta_1+ \beta_0)^\top \Sigma_X (\beta_1+\beta_0) /\sigma_{\mathrm {CL}}^2 \geq 1 $ with the strict inequality holding unless $ \beta_1+\beta_0= 0 $. 
Thus,  ${\cal T}_{\mathrm {CL}}$ has a
guaranteed efficiency gain over  ${\cal T}_{\mathrm L}$ under simple randomization. 

Under covariate-adaptive randomization satisfying (D$^\dag$) with $\nu < \pi (1-\pi )$, Theorem 2(b) shows that ${\cal T}_{\mathrm L}$ is not valid but conservative as $\sigma^2_{\mathrm L} (\nu )< \sigma^2_{\mathrm L}$ unless $ E(O_{i1} | Z_i) + 
E(O_{i0} | Z_i)=0$ almost surely under $H_0$, which holds under some extreme scenarios, e.g., $Z$ used for randomization is independent of the outcome. 
This conservativeness can be corrected by a multiplication factor $\hat{r}(\nu) \xrightarrow{p} \sigma_{\mathrm L} / \sigma_{\mathrm L} (\nu )$  under $H_0$.
The resulting $ \hat{r}(\nu ) {\cal T}_{\mathrm L} $ is  
the modified log-rank test in \cite{Ye:2020survival}, which is valid and always more powerful than ${\cal T}_{\mathrm L}$.
Under the local alternative  specified in Theorem 1(c) with  $\pi c_1 - (1-\pi )c_0 \neq 0$,
Pitman's  asymptotic relative efficiency of $\hat{r}(\nu ){\cal T}_{\mathrm L}  $ with respect to  $ {\cal T}_{\mathrm {CL}}  $ in (\ref{eq: TCL})  is  
$ \sigma_{\mathrm L}^2(\nu ) /\sigma_{\mathrm {CL}}^2 
= 1+ (\beta_1+\beta_0)^\top \left[ \pi (1-\pi) E\{   \var (X_i \mid Z_i)\} + \nu \var\{   E (X_i \mid Z_i)\}  \right] (\beta_1+\beta_0) /\sigma^2_{\mathrm {CL}} \geq 1
$ 
with the strict inequality holding unless $\beta_1+\beta_0 =0$, e.g., $X_i$ is uncorrelated with $O_{ij}$,  or $ \nu=0$ and $E \{\cov (X_i, O_{i1} \mid Z_i )
+ \cov (X_i , O_{i0} \mid  Z_i) \} = 0$, e.g., covariates in $X_i$ but not in $Z_i$ are uncorrelated with $O_{ij}$ conditioned on $Z_i$. 
Hence, the adjusted ${\cal T}_{\mathrm {CL}}$ has a guaranteed efficiency gain over both the log-rank test ${\cal T}_{\mathrm L}$ and modified  log-rank test $\hat{r}(\nu ){\cal T}_{\mathrm L}  $ 
under any covariate-adaptive randomization schemes satisfying (D) and (D$^\dag$). 

Note that Pocock and Simon's minimization satisfies (D) but not necessarily (D$^\dag$) as $I_i$'s are correlated across strata. 
Hence, under Pocock and Simon's minimization,  Theorem 2 is not applicable  and 
${\cal T}_{\mathrm L}$ may not be valid, whereas 
${\cal T}_{\mathrm {CL}}$ is valid according to Theorem 1,
another advantage of covariate adjustment.

$\hat U_{\mathrm {CL}}$ in the numerator of (\ref{eq: TCL}) is the same as the  augmented score in \cite{Lu:2008}, which shares the same idea as those in \cite{Tsiatis:2008aa} and \cite{zhang2008improving} for non-censored data. 
However, the denominator $\hat\sigma_{\rm CL}$ in (\ref{eq: TCL}) is different from that used by \cite{Lu:2008}.  
The key difference between our result on guaranteed efficiency gain and the result  in \cite{Lu:2008} is, our result is obtained under covariate-adaptive randomization and an alternative hypothesis without any specific model on the distribution of $T_j$ or $C_j$, whereas the result in \cite{Lu:2008} is for simple randomization and  an alternative under 
a correctly specified Cox proportional hazards model for $T_j$.

After testing $H_0$, it is often of interest to  estimate and construct a confidence interval for an effect size
\citep{Lu:2008, parast2014landmark, zhang2015robust,diaz2019improved}. A commonly considered effect size is the hazard ratio $e^{ \theta }$ under the Cox proportional hazards model $\lambda_1(t)  = \lambda_0 (t) e^{\theta} $.  Note that the hazard ratio $e^{ \theta }$ is interpretable only when the Cox proportional hazards model is correctly specified. Thus,  in the rest of this section we consider covariate-adjusted  estimation and confidence interval for $\theta$, assuming  $\lambda_1(t)  = \lambda_0 (t) e^{\theta} $. 

Without using any covariate, the score from the partial likelihood under model  $\lambda_1(t)  = \lambda_0 (t) e^{\theta} $ is \vspace{-1mm}
\begin{align*}
\widehat U_{\mathrm L} (\vartheta ) = \frac1n  \sum_{i=1}^n \int_0^\tau \bigg\{ I _i - \frac{e^\vartheta \bar Y_1(t) }{e^\vartheta \bar Y_1(t) + \bar Y_0 (t) }  \bigg\} dN_i(t). 
\end{align*}
The maximum partial likelihood estimator $\widehat\theta_{\mathrm L} $ of $\theta$ is 
a solution to $\widehat U_{\mathrm L} ( \vartheta ) = 0$. Using the  idea  in (\ref{CL1}) with $X_i$ containing all levels of $ Z_i $  used in  covariate-adaptive
randomization,
our covariate-adjusted score  is \vspace{-1mm}
\begin{align*}
\widehat U_{\mathrm {CL}} (\vartheta ) & = \widehat U_{\mathrm L} (\vartheta ) -  \frac1n \sum_{i=1}^n \{ I_i (X_i - \bar X)^\top \widehat \beta_1(\widehat \theta_{\mathrm L}) - (1-I_i) (X_i - \bar X)^\top \widehat \beta_0(\widehat \theta_{\mathrm L})  \} , 
\end{align*}
where, for $j=0,1$, $		\widehat \beta_j (\vartheta ) $ is equal to 
$\widehat\beta_j$ in (\ref{hatbeta}) with 
$ \widehat O_{ij}$ replaced by   
\begin{align*}
\widehat O_{ij}(\vartheta)   = \int_0^\tau  \frac{\{ e^{\vartheta} \bar Y_1(t) \}^{(1-j)} \{ \bar Y_0(t)\}^{j}  }{e^{\vartheta}  \bar Y_1(t) + \bar Y_0 (t)}  \left\{ dN_{ij}(t) - \frac{  Y_{ij}(t) e^{j \vartheta}  d \bar N(t)}{e^{\vartheta}  \bar Y_1(t) + \bar Y_0 (t) }\right\} .
\end{align*}
Solving $	\widehat U_{\mathrm {CL}} (\vartheta ) = 0$  gives the covariate-adjusted estimator $\widehat \theta_{\mathrm {CL}}$. As $	\widehat U_{\mathrm {CL}} (\theta )$ has reduced variability compared to $\widehat U_{\mathrm L} (\theta ) $, and $ \partial \widehat U_{\mathrm {CL}} (\vartheta) / \partial \vartheta   = \partial \widehat U_{\mathrm L} (\vartheta ) / \partial \vartheta$, by a standard argument for M-estimators, $\widehat \theta_{\mathrm {CL}}$ is guaranteed to have smaller variance than $\widehat \theta_{\mathrm L}$. 
It is established in Section S2.2 of   Supplementary Material that 
$\sqrt{n} ( \widehat \theta_{\mathrm {CL}}- \theta ) \xrightarrow{d} N(0, \sigma^2(\theta ))$ under any covariate-adaptive randomization satisfying (D),  with 
$	\sigma^2 (\theta) $ given in Theorem S2 in  Supplementary Material. 
An asymptotic confidence interval for $\theta$ can be obtained based on this result
and a consistent estimator of $\sigma^2(\theta )$ given by  
$[ g(\hat\theta_{\mathrm {CL}}) - \pi(1-\pi) \{\hat \beta_1(\hat\theta_{\mathrm {L}}) + \hat\beta_0(\hat\theta_{\mathrm {L}}) \}^\top \hat \Sigma_X \{\hat\beta_1(\hat\theta_{\mathrm {L}}) +\hat\beta_0(\hat\theta_{\mathrm {L}}) \} ]/ \{g( \hat\theta_{\mathrm {CL}})\}^2$, where $g(\vartheta ) = - \partial \widehat U_{\mathrm L} (\vartheta ) / \partial \vartheta$.

\section{Covariate-Adjusted Stratified Log-Rank Test}

The stratified log-rank test \citep{Peto:1976aa} is a weighted average of the stratum-specific log-rank test statistics  with finitely many strata constructed using a discrete baseline covariate. 
{We consider stratification with  all levels of 
$Z_i$. Results can be obtained similarly for stratifying on more  levels than those of $Z_i$ or fewer levels than those of $Z_i$ with levels of $Z_i$ not used in stratification included in $X_i$}. Here, we remove the part of $X_i$ that can be linearly represented by $Z_i$ and still denote the remaining as $X_i$. As such, it is reasonable to assume that  $E\{\var (X_i \mid Z_i)\}$  is positive definite. 

The stratified log-rank test using levels of $Z_i$ as strata is 
\begin{equation}
{\cal T}_{\mathrm {SL}} = \sqrt{n} \, \hat U_{\mathrm {SL}}  / \, \hat\sigma_{\mathrm {SL}} , \label{SL}
\end{equation} 
where
$$
\hat	U_{\mathrm {SL}}  = \frac{1}{n}  \sum_z \sum_{i: Z_i=z} \int_{0}^{\tau} \left\{I_i-\frac{\bar{Y}_{z1}(t)}{\bar Y_z(t)}\right\}dN_i(t),  \quad 
\hat \sigma^2_{\mathrm {SL}} =  \frac{1}{n} \sum_z \sum_{i: Z_i=z}\int_0^\tau \frac{\bar Y_{z1}(t)\bar Y_{z0} (t)}{ \bar Y_z (t) ^2}dN_i(t), 
$$
$ \bar Y_{z1} (t)=  \sum_{i: Z_i = z} I_i Y_i(t)/n$,  $\bar Y_{z0} (t)=  \sum_{i: Z_i =z} (1-I_i) Y_i(t)/n$, and $\bar Y_z (t) = \bar Y_{z1}(t)+\bar Y_{z0}(t)$.

With stratification,  ${\cal T}_{\rm SL}  $ in (\ref{SL}) actually tests the null hypothesis 
$\tilde H_0: \lambda_1 (t \mid z) = \lambda_0(t\mid z)$ for all $(t,z)$, 
where $\lambda_j(t\mid z) $ is the hazard function of $T_j$ conditional on $Z=z$. 
Hypothesis $\tilde H_0$ may be stronger than $H_0: \lambda_1(t) = \lambda_0(t)$ for all $t$, the null hypothesis for unstratified log-rank test ${\cal T}_{\rm L}  $ and its adjustment ${\cal T}_{\rm CL}  $ considered in \S2-3. In some scenarios, $\tilde H_0=H_0$; for example, when (TR) holds with $Z \subset V$. 

To further adjust for baseline covariate $X_i$, 
we still linearize $	\hat	U_{\mathrm {SL}} $ as follows \citep{Ye:2020survival}, 
\begin{align*}
\hat  U_{\mathrm {SL}} &= \frac{1}{{n}}  \sum_z \sum_{i: Z_i= z}   \left\{  I_i \olim_{zi1}  - (1-I_i)  \olim_{zi0}  \right\} + o_p(n^{-1/2}),\nonumber  
\end{align*}
where 
$$ 
\olim_{zij} 
=\int_0^\tau \{ 1-\mu_z(t)\}^j \{\mu_z(t)\}^{1-j}  \{dN_{ij}(t)-Y_{ij}(t)  p_z(t) dt\},  \quad 	j=0,1, $$ 
$p_z(t) dt = {E}\{d N_i(t) \mid Z_i= z \} /E\{Y_i(t) \mid Z_i= z\} $, and $\mu_z(t)=  E (I_i \mid Y_i(t)= 1, Z_i=z)  $. Following the same idea in Section 3, 
we  apply the generalized regression adjustment by using   
\begin{align*}
\ohat_{zij}   = \int_{0}^\tau \frac{ \bar Y_{z(1-j)} (t)}{\bar Y_z(t)} \left\{  dN_{ij} (t) 
- Y_{ij} (t)   \frac{d\bar N_z(t) }{ \bar Y_z (t)} \right\},  \quad j=0,1, 
\end{align*}
as derived outcomes,  
where $\bar N_z(t)= \sum_{i: Z_i=z} N_i(t)/n$. 
The resulting covariate-adjusted version of $\hat  U_{\mathrm {SL}}$ is
\begin{align*}
\hat  U_{\mathrm {CSL}} 
&=  \frac{1}{n} \sum_z \sum_{i: Z_i= z}   \left[ I_i \{ \ohat_{zi1}  -  (X_i- \bar X_z)^\top \hat\gamma_1  \} - (1-I_i)  \{ \ohat_{zi0}   - ( X_i- \bar X_z)^\top \hat \gamma_0  \}  \right]\\
&=  \hat U_{\mathrm {SL}} - \frac{1}{n} \sum_z \sum_{i: Z_i= z}   \left\{ I_i  (X_i- \bar X_z)^\top \hat\gamma_1  - (1-I_i)    ( X_i- \bar X_z)^\top \hat \gamma_0    \right\} , 
\end{align*}
where the last equality is from the algebraic identify $\hat U_{\mathrm {SL}}  = n^{-1}\sum_z \sum_{i: Z_i=z} \{ I_i \ohat_{zi1}  - (1-I_i) \ohat_{zi0}  \}, $ $\bar X_z$ is the sample mean of $X_i$'s with $Z_i=z$, 
\begin{align*}
\hat\gamma_j =    \bigg\{ \sum_z \sum_{i: I_i= j, Z_i=z} (X_i -  \bar X_{zj}  ) (X_i -  \bar X_{zj}  )^\top   \bigg\}^{-1}    \sum_z \sum_{i: I_i= j, Z_i=z} (X_i -  \bar X_{zj}  ) \ohat_{zij}  
\end{align*}
converging  to a limit value  $\gamma_j $ in probability (Lemma S1 of Supplementary Material),  and 
$\bar X_{zj}$ is the sample mean of $X_i$'s with $Z_i=z$ and treatment $j$, $j=0,1$.
Our proposed covariate-adjusted stratified log-rank test is 
\begin{equation}
{\cal T}_{\mathrm {CSL}}= \sqrt{n} \, \hat  U_{\mathrm {CSL}}/ \, \hat\sigma_{\mathrm {CSL}}, \label{CSL}
\end{equation}
where
$ \hat\sigma_{\mathrm {CSL}}^2 = \hat\sigma^2_{\mathrm {SL}} - \pi (1-\pi) 
(\hat\gamma_1+\hat\gamma_0)^\top  \big\{ \sum_z (n_z/n )\hat \Sigma_{X|z} \big\} (\hat\gamma_1+\hat\gamma_0)$
and	  $\hat \Sigma_{X|z} $ is the sample covariance matrix of $X_i$'s  within stratum $z$.

The following theorem establishes the asymptotic properties of the stratified log-rank test ${\cal T}_{\mathrm {SL}}$ and covariate-adjusted stratified log-rank test ${\cal T}_{\mathrm {CSL}}$. 

\begin{theorem} \label{theo: stratified logrank} 	Assume   $C_I \perp (T_I, W) \mid (I, Z)$ or (CR)-(TR) with $Z \subset V$. 
Assume also {(D)}.  
Then, the following results hold regardless of which covariate-adaptive randomization is applied. \vspace{-2mm}
\begin{itemize}
\item[(a)]    Under the null $\tilde H_0$ or  alternative hypothesis,
$ \sqrt{n}  \{ \hat  U_{\mathrm {CSL}} -  \sum_z (n_{z1}\theta_{z1} - n_{z0}\theta_{z0})/n  \}  \xrightarrow{d} N\left(0,\sigma_{\mathrm {CSL}}^2 \right) $, 
and the same result holds with $\hat  U_{\mathrm {CSL}}$ and $\sigma_{\mathrm {CSL}}^2 $ replaced by $\hat  U_{\mathrm {SL}} $ and $\sigma_{\mathrm {SL}}^2 $, respectively, 
where 
$\theta_{zj} = E(\olim_{zij} \mid Z_i = z)$, $n_{zj} =$ the number of patients with  treatment $j$ in stratum $z$,  $j=0,1$,  
$	\sigma_{\mathrm {CSL}}^2 
= \sigma_{\mathrm {SL}}^2 - \pi(1-\pi) (\gamma_1 + \gamma_0)^\top  E\{ \var (X_i \mid Z_i)\}(\gamma_1 + \gamma_0) $, and 
$		\sigma_{\mathrm {SL}}^2 = \sum_z \P (Z_i=z)  \{ \pi \var (\olim_{zi1} \mid Z_i =z) + (1-\pi )  \var (\olim_{zi0} \mid Z_i =z)\} $. 
\item[(b)] 	  Under the null  hypothesis $\tilde H_0$, \ $\theta_{z1} = \theta_{z0} =0$ \ for any $z$, \
$\hat\sigma^2_{\mathrm {SL}} \xrightarrow{p} \sigma^2_{\mathrm {SL}}$, \
$\hat\sigma^2_{\mathrm {CSL}}\xrightarrow{p} \sigma^2_{\mathrm {CSL}}$, \
${\cal T}_{\mathrm {SL}} \xrightarrow{d} N(0,1)$,  \ and \
${\cal T}_{\mathrm {CSL}}\xrightarrow{d} N(0,1)$, i.e.,  both ${\cal T}_{\mathrm {SL}} $ and ${\cal T}_{\mathrm {CSL}}$
are valid  for testing null hypothesis $\tilde H_0$. 
\item[(c)]  	Under the local alternative hypothesis that 	$\theta_{zj} = c_{zj}n^{-1/2}$ with $c_{zj}$'s not depending on $n$  and that 
$\lambda_1(t\mid z) / \lambda_0(t\mid z) $ is bounded and $\to 1$  for every $t$ and $z$,  
$ {\cal T}_{\mathrm {CSL}}\xrightarrow{d}  N(  \sum_z   \P (Z=z) \{ \pi c_{z1}- (1-\pi )c_{z0}\}/ \sigma_{\mathrm {CSL}}, \ 1) $,
and the same result holds with ${\cal T}_{\mathrm {CSL}}$ and $\sigma_{\mathrm {CSL}}$ replaced by  ${\cal T}_{\mathrm {SL}}$ and $\sigma_{\mathrm {SL}}$, respectively. 
\end{itemize}
\end{theorem}

Like ${\cal T}_{\mathrm {CL}}$ in \eqref{eq: TCL}, 
both ${\cal T}_{\mathrm {SL}}$ in (\ref{SL}) and  
${\cal T}_{\mathrm {CSL}}$ in (\ref{CSL}) are applicable to all covariate-adaptive randomization schemes with universal formulas, i.e., they achieve the universal applicability. 	In terms of Pitman's asymptotic efficiency under the local alternative specified in Theorem 3(c), ${\cal T}_{\mathrm {CSL}}$ is always more efficient than ${\cal T}_{\mathrm {SL}}$,  since 
$ \sigma_{\mathrm {CSL}}^2  \leq  \sigma_{\mathrm {SL}}^2$with the strict inequality holding  unless $ \gamma_1+\gamma_0= 0 $.

The condition $C_I \perp (T_I, W) \mid (I, Z)$ in Theorem 3 for stratified log-rank test and its adjustment is weaker than condition (C) in Theorem 1 for unstratified log-rank test. However, the hypothesis $\tilde H_0$ may be stronger than $H_0$. 

Is ${\cal T}_{\mathrm {SL}}$ or ${\cal T}_{\mathrm {CSL}}$ more efficient than the unstratified log-rank test ${\cal T}_{\mathrm L}$? The answer is not definite because, first of all, 
the null hypotheses $\tilde H_0$ and $H_0$ may be different as we discussed earlier, and secondly, even if $\tilde H_0 =H_0$, under the alternative 
the asymptotic mean $(n_1 \theta_1 - n_0 \theta_0) /n $ of $\hat U_{\mathrm L}$ may not be comparable with  the asymptotic mean 
$\sum_z (n_{z1}\theta_{z1}-n_{z0}\theta_{z0})/n$ of $\hat U_{\mathrm {SL}}$ or $\hat U_{\mathrm {CSL}}$.  
In fact, the indefiniteness of relative efficiency between the stratified and unstratified log-rank tests is a standing problem  in the literature. 

There is also no definite answer when comparing the efficiency of ${\cal T}_{\mathrm {CL}}$ and the stratified ${\cal T}_{\mathrm {CSL}}$.

Similar to the discussion in the end of Section 3, 
after testing hypothesis $\tilde H_0$, we can obtain a 
covariate-adjusted confidence interval for the effect size  $\theta$ 
under a stratified   Cox proportional hazards model $\lambda_{1z}(t)  = \lambda_{0z} (t) e^\theta $ for every $z$. The details are in {Section S2.3} of Supplementary Material.

\section{Simulations}

To supplement theory and examine finite sample type I error and power of tests 
${\cal T}_{\mathrm L}$, ${\cal T}_{\mathrm {CL}}$, ${\cal T}_{\mathrm {SL}}$, and ${\cal T}_{\mathrm {CSL}}$, we carry out a simulation study under the following four cases/models.  
\begin{description}
\item Case I: The conditional  hazard  follows a Cox model, $ \lambda_j (t \mid W) = (\log 2) \exp ( - \theta j + \eta^\top W)$ for $j=0,1$, where $\theta$ denotes a scalar parameter, 
$\eta = (0.5,0.5,0.5)^\top $, and
$W$ is a 3-dimensional covariate vector following the 3-dimensional standard  normal distribution. The censoring variables $C_0$ and $C_1$  follow uniform distribution on interval  $(10,40)$ and are independent of $W$.
\item Case II: The conditional hazard is the same as that in case I.  Conditional on $W$ and treatment assignment $j$,  $C_j - (3- 3j)$  follows a standard exponential distribution. 
\item Case III: $T_j = \exp ( - \theta j + \eta^\top W)+ {\cal E}$, $j=0,1$,  
where $\theta$, $\eta $, and	$W$ are the same as those in case I, and 
${\cal E}$ is a random variable independent of  $(C_1,C_0,W)$ and has the standard exponential distribution.
The setting for censoring is the same as that in case I. 
\item Case IV:  The models  for $T_j$'s and $C_j$'s are the same as that in case III and
case II, respectively. 
\end{description}
In this simulation, the significance level  $\alpha=5\%$, the target treatment assignment proportion  $\pi = 0.5$, the overall sample size $n=500$,  and 
the null hypothesis $H_0: \theta =0$. 
Three randomization schemes are considered, simple randomization, stratified permuted block with block size 4 and levels of $Z$ as strata, and 
Pocock and Simon's minimization assigning a patient with probability 0.8 to the preferred arm minimizing the sum of balance scores over marginal levels of $Z$, where $Z$ is the 2-dimensional vector whose first component is a two-level discretized first component of $W$ and second component is  a three-level 
discretized second component of  $W$. 
For stratified log-rank tests, levels of $Z$ are used as  strata.  For covariate adjustment, $X $ is the vector containing $Z$ and the third component of $W$  for ${\cal T}_{\mathrm {CL}}$,
and $X$ is the third component of $W$ for ${\cal T}_{\mathrm {CSL}}$. 

Based on 10,000 simulations, type I error rates for four tests under four cases and three randomization schemes are shown in Table 1. 
The results agree with our theory.
For ${\cal T}_{\mathrm {CL}}$, ${\cal T}_{\mathrm {SL}}$, and ${\cal T}_{\mathrm {CSL}}$, 
there is no substantial difference among the three randomization schemes.  The log-rank test ${\cal T}_{\mathrm L}$ preserves 5\% rate under simple randomization, but it is conservative under stratified permuted block and minimization.

Based on 10,000 simulations, 
power curves of four tests for $\theta $ ranging from 0 to 0.6, under four cases and stratified permuted block randomization are plotted in Figure 1. Similar figures for simple randomization and minimization are given in  Supplementary Material. 
In all cases, the power curves of covariate-adjusted tests ${\cal T}_{\mathrm {CL}}$ and 
${\cal T}_{\mathrm {CSL}}$ are better than those of unadjusted tests  ${\cal T}_{\mathrm L}$ and 
${\cal T}_{\mathrm {SL}}$,  especially the benchmark ${\cal T}_{\mathrm L}$. 
Under Cox's model, ${\cal T}_{\mathrm {CSL}}$ is better than ${\cal T}_{\mathrm {CL}}$,
but not necessarily under non-Cox model. 
The stratified 
${\cal T}_{\mathrm {SL}}$ is mostly better than the unstratified ${\cal T}_{\mathrm L}$, 
but unlike ${\cal T}_{\mathrm {CL}}$ and 
${\cal T}_{\mathrm {CSL}}$, there is no guaranteed efficiency gain, e.g., case III when $\theta > 0.4$. 
The difference in censoring model also has some effect.   

More simulation results can be found in Supplementary Material. 

\section{A Real Data Application}
We apply four tests ${\cal T}_{\mathrm L}$, ${\cal T}_{\mathrm {CL}}$, ${\cal T}_{\mathrm {SL}}$, and ${\cal T}_{\mathrm {CSL}}$   to the data  from the AIDS Clinical Trials Group Study 175 (ACTG 175), a randomized controlled trial evaluating  antiretroviral treatments in adults infected with human immunodeficiency virus type 1  whose CD4 cell counts were from 200 to 500 per cubic millimeter \citep{Hammer:1996wk}. The primary endpoint was time to a composite event defined as
a $\geq 50$\% decline in  CD4 cell count, an AIDS-defining event, or death. 
Stratified permuted block randomization with equal allocation was applied
with covariate $Z$ having three levels related with the length of prior antiretroviral therapy: $Z=1$, 2, and 3 
representing 0 week,  between 1 to $52$ weeks, and  more than $52 $ weeks of prior antiretroviral  therapy, respectively. 
The dataset is publically available in the \textsf{R} package \textsf{speff2trial}.

We focus on the comparison of treatment 0 (zidovudine) versus treatment 1 (didanosine).  For stratified log-rank test 
${\cal T}_{\mathrm {SL}}$,  the three-level $Z$ is used as the stratification variable. 
For covariate adjustment,  two additional prognostic baseline covariates are considered as $X$, the baseline CD4 cell count and number of days  receiving antiretroviral therapy prior to treatment. 
In addition to testing treatment effect for all patients, a sub-group analysis with $Z$-strata as sub-groups is also of interest  because responses to antiretroviral therapy may vary according to the extent of prior drug exposure.
Within each sub-group defined by $Z$, the stratified tests become the same as their unstratified counterparts, and thus we only apply tests ${\cal T}_{\mathrm L}$ and ${\cal T}_{\mathrm {CL}}$ in the sub-group analysis. 

Table \ref{tb: data} reports the number of patients, numerator and denominator of  
each test, and p-value 
for testing with all patients or with a sub-group. 
The effect of covariate adjustment is clear:  for the  covariate-adjusted tests, 
the standard errors $\hat\sigma_{\mathrm {CL}}$ and $\hat\sigma_{\mathrm {CSL}}$ are smaller than $\hat\sigma_{\mathrm L}$  and $\hat\sigma_{\mathrm {SL}}$ in all analyses. 


For the analysis based on all patients, all four tests  significantly reject the null hypothesis $H_0$ of no treatment effect. 
In sub-group analysis, the p-values are adjusted using Bonferroni's correction to control for the family-wise error rate. From Table \ref{tb: data}, p-values in sub-group analysis are substantially larger than those in the analysis of all patients, because of reduced sample sizes as well as Bonferroni's correction. 
The empirical result in this example illustrates the benefit  of covariate-adjustment in testing when sample size is not very large. 
Using the adjusted log-rank test ${\cal T}_{\mathrm {CL}}$, together with  the estimated effect size and its standard error  shown in Table 2,
we can conclude the superiority of treatment 1  for both $Z=1$ and $Z=3$, which is consistent with the evidence in \cite{Hammer:1996wk}. 

\section*{Acknowledgement}

We would like to thank all reviewers for useful comments and suggestions. Dr. Jun Shao was supported
by grants from the National Natural Science Foundation of China and U.S. National Science Foundation.

\section*{Supplementary Material}

The supplementary material contains   all technical proofs and some additional results.

\bibliographystyle{apalike}
\bibliography{reference}

\begin{figure}[b]
\centering
\caption{Power curves based on 10,000 simulations} 	\includegraphics[scale=0.5]{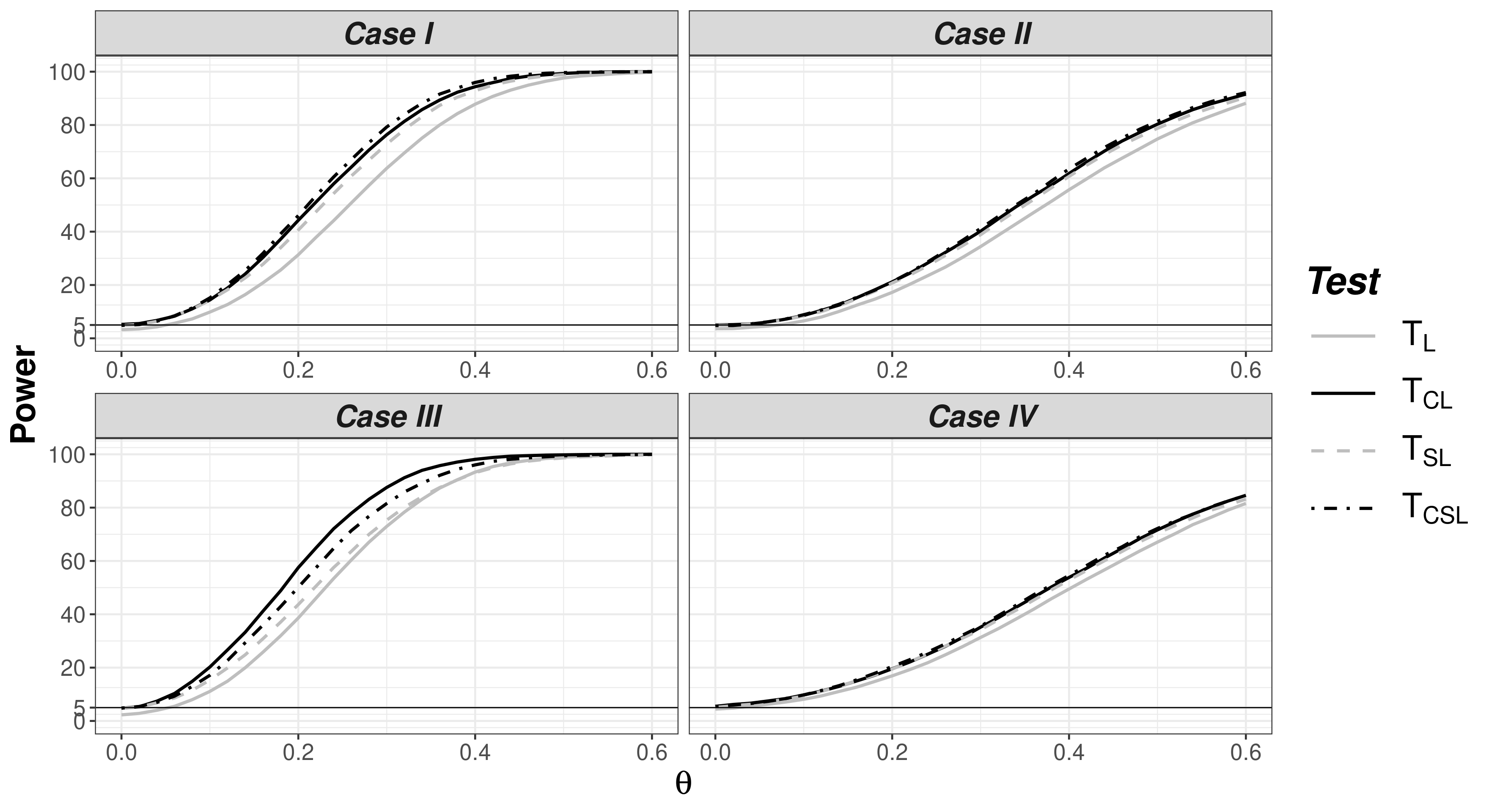}
\end{figure}

\newpage

\begin{table}[]	
\begin{center}
\caption{Type I errors (in \%) based on 10,000 simulations}
{\footnotesize		\begin{tabular}{clccccc}				\hline \\[-1.5ex]
		Case		&  \multicolumn{1}{c}{Randomization}     &   & ${\cal T}_{\mathrm L}$    & ${\cal T}_{\mathrm {CL}}$ & ${\cal T}_{\mathrm {SL}}$ & ${\cal T}_{\mathrm {CSL}}$           \\ [1ex]\hline
		I & simple & & 4.91 & 5.16 & 4.86 & 4.78   \\	
		& permuted block & & 3.25 & 5.22 & 4.80 & 4.85\\	
		& minimization & & 3.40 & 5.43 & 5.02 & 5.23  \\	
		II & simple & & 5.39 & 5.14 & 5.00 & 4.97\\	
		& permuted block & & 3.59 & 5.03 & 4.94 & 4.82 \\	
		& minimization & & 4.01 & 5.23 & 5.11 & 5.28  \\	
		III & simple & & 5.07 & 5.43 & 5.27 & 5.16 \\	
		& permuted block & & 2.29 & 4.79 & 4.76 & 4.82 \\	
		& minimization & &2.88 & 5.43 & 5.23 & 5.52  \\	
		IV & simple & & 5.41 & 5.30 & 5.39 & 5.21 \\	
		& permuted block & & 4.44 & 5.48 & 5.10 & 5.49 \\	
		& minimization & & 4.21 & 5.18 & 5.04 & 5.06  \\	
		[0.5ex] \hline \\[-1.5ex]
\end{tabular}}
\end{center}
\end{table}

\begin{table}[]	
\begin{center}
\caption{Statistics for the ACTG 175 example \label{tb: data}}
{\footnotesize		\begin{tabular}{lrrrrr}				\hline \\[-1.5ex]
		& & & \multicolumn{3}{c}{Sub-group} \\
		& \multicolumn{1}{c}{All patients}     &  & $Z=1$    & $Z=2$   & $Z=3$             \\[0.5ex] \hline \\[-1.5ex]
		Number of patients & 1,093 && 461 & 198 & 434 \\ [0.5ex]
		Log-rank 
		&                  &  &        &       &                  \\
		\quad $\sqrt{n} \hat U_{\mathrm L}$              & -1.223           &  &  -0.542 &  -0.144 & -1.292           \\
		\quad \quad  $ \hat \sigma_{\mathrm L}$            & 0.265           &  & 0.235  & 0.270 & 0.290        \\
		\quad 	 p-value (adjusted for sub-group analysis)& \textless{}0.001 &  & 0.064  & 1     & \textless{}0.001 \\
		\quad Estimated $\theta $ & -0.528 && -0.455 & -0.140&-0.740   \\
		\quad Standard error of the estimated $\theta $ & 0.116 && 0.199&  0.263&  0.171 \\ [0.5ex]
		Covariate-adjusted log-rank 
		&                  &  &        &       &                  \\
		\quad $\sqrt{n} \hat U_{\mathrm {CL}}$           &  -1.273       &  & -0.553 &  -0.129 &  -1.382           \\
		\quad	\quad   $ \hat \sigma_{\mathrm {CL}}$             & 0.257            &  & 0.230  & 0.265 & 0.282         \\
		\quad	p-value (adjusted for sub-group analysis) & \textless{}0.001 &  & 0.049  & 1     & \textless{}0.001 \\
		\quad Estimated $\theta $  & -0.550 && -0.464 & -0.127 &-0.793   \\
		\quad Standard error  of the estimated $\theta $ &0.113 && 0.195&  0.257&  0.166 \\ [0.5ex] 
		Stratified log-rank 
		&                  &  &        &       &                  \\
		\quad $\sqrt{n} \hat U_{\mathrm {SL}}$                & -1.228           &  &        &       &                  \\
		\quad   \quad $\hat \sigma_{\mathrm {SL}}$        & 0.264           &  &        &       &                  \\
		\quad	p-value            & \textless{}0.001 &  &        &       &                  \\
		\quad Estimated $\theta $  & -0.531 &&  \\
		\quad Standard error  of the estimated $\theta $ &0.116 && \\ [0.5ex]  
		Covariate-adjusted stratified log-rank 
		&                  &  &        &       &                  \\
		\quad	$\sqrt{n} \hat U_{\mathrm {CSL}}$       & -1.284           &  &        &       &                  \\
		\quad   \quad $\hat \sigma_{\mathrm {CSL}}$              & 0.258         &  &        &       &                  \\
		\quad	p-value            & \textless{}0.001 &  &        &       &                 \\
		\quad Estimated $\theta $  &-0.556 &&  \\
		\quad Standard error  of the estimated $\theta $ &0.113 && \\ 
		\hline 
		\multicolumn{6}{l}{$\theta$ respectively denotes log hazard ratio for all patients and for each subgroup.} \\
\end{tabular}}
\end{center}
\end{table}

\clearpage


\setcounter{equation}{0}
\setcounter{table}{0}
\setcounter{lemma}{0}
\setcounter{section}{0}
\setcounter{figure}{0}
\setcounter{theorem}{0}
\renewcommand{\theequation}{S\arabic{equation}}
\renewcommand{\thelemma}{S\arabic{lemma}}
\renewcommand{\thetheorem}{S\arabic{theorem}}
\renewcommand{\thefigure}{S\arabic{figure}}
\renewcommand{\thetable}{S\arabic{table}}

\begin{center}
	{\sffamily\bfseries\LARGE
		Supplementary Materials
	}
\end{center}

\section{Additional Simulations}

\subsection{Additional simulations with $n=500$ under Case I-IV }

Based on 10,000 simulations, power curves of four tests for $\theta$ ranging from 0 to 0.6 under simple randomization and minimization are  in Figure \ref{fig,n=500}.

\begin{figure}[h]
	\centering
	\includegraphics[scale=0.5]{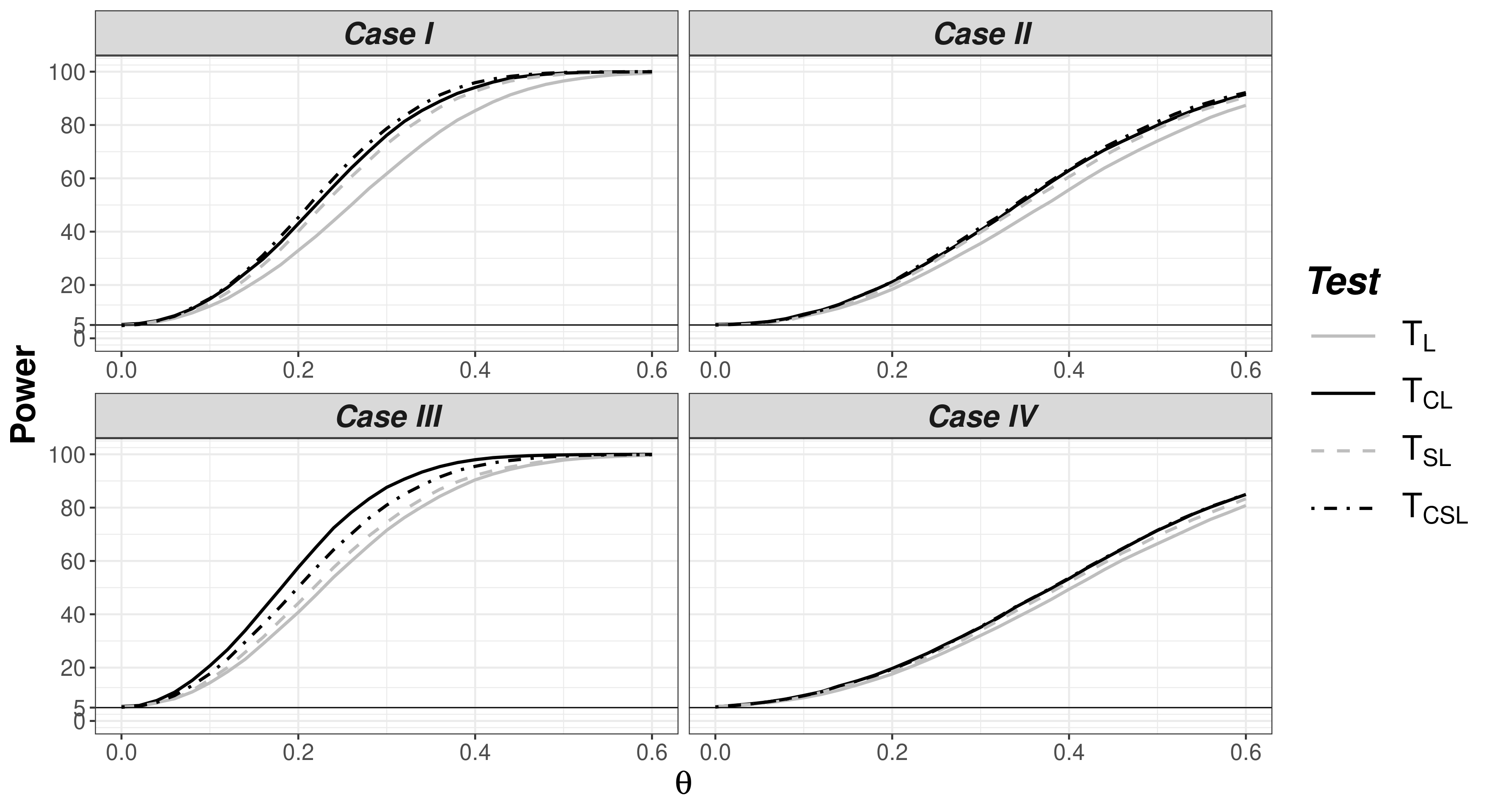}
	\includegraphics[scale=0.5]{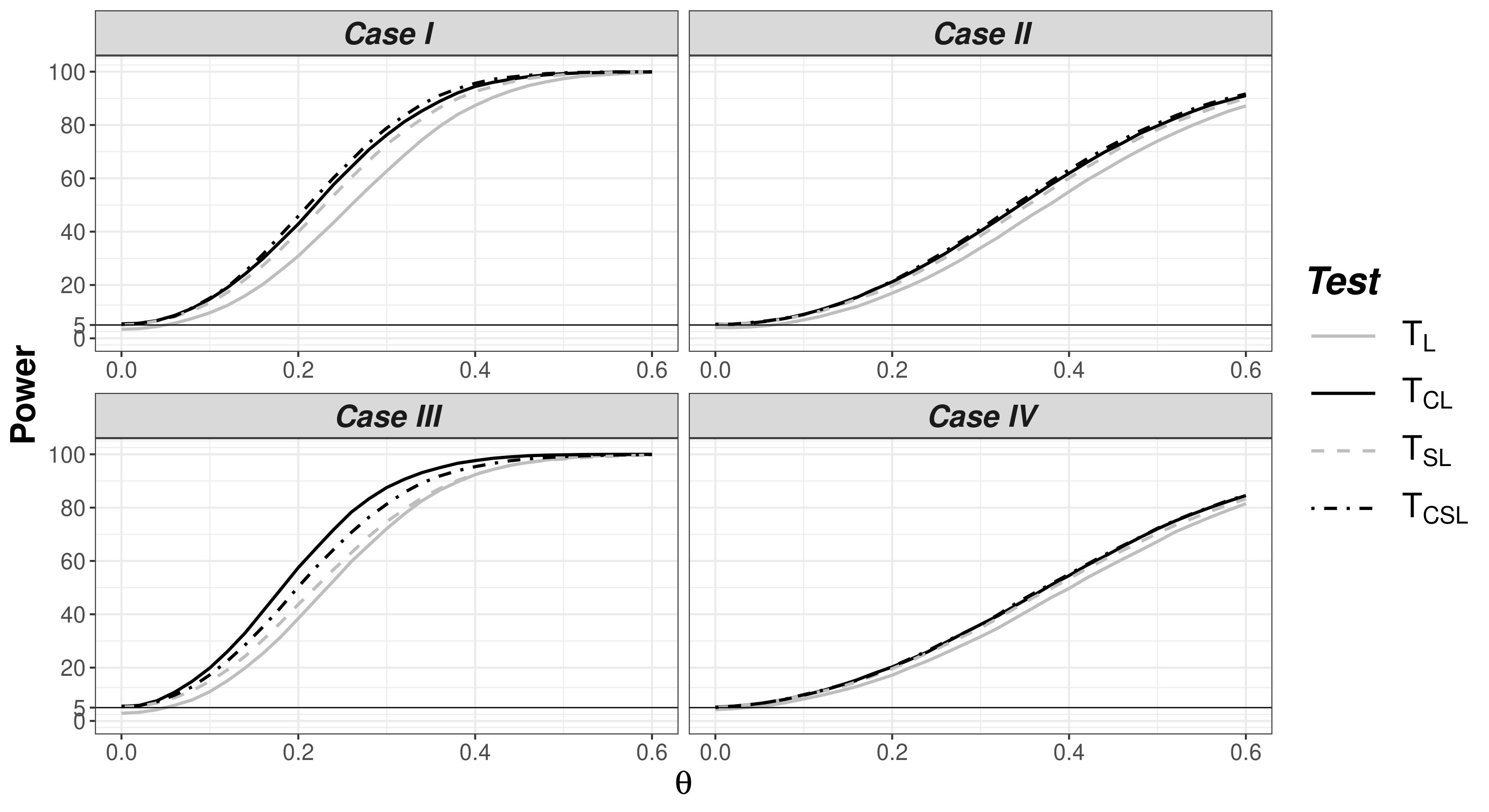}
	\caption{Power curves based on 10,000 simulations with $n=500$ under simple randomization (top) and minimization (bottom). \label{fig,n=500}} 
\end{figure}

\subsection{Simulations with $n=200$ under Case I-IV }
The simulation setting is the same as the simulations in the main article, except that $n=200$ and $Z$ is the 2-dimensional vector whose first component is a two-level discretized first component of $W$ and second component is  a two-level 
discretized second component of  $W$. Thus,  the average sample size in each treatment and $Z$-level combination is $200/(2\times 4) = 25$.  Type I error rates for four tests under four cases and three randomization schemes are shown in Table \ref{tab: type I error n=200}. Power curves under three randomization schemes are in Figure \ref{fig: n=200}. From Table \ref{tab: type I error n=200}, we see that with a smaller sample size, the type I error rates of the two covariate-adjusted tests  $\mathcal{T}_{\mathrm{CL}}$ and $\mathcal{T}_{\mathrm{CSL}}$ can be slightly inflated but the inflation is not too severe. Otherwise, the results with $n=200$ are similar to the results with $n=500$.

\begin{table}[h]	
	\begin{center}
		\caption{Type I errors (in \%) based on 10,000 simulations with $n=200$ \label{tab: type I error n=200}}
		\begin{tabular}{clccccc}				\hline \\[-1.5ex]
			Case		&  \multicolumn{1}{c}{Randomization}       & ${\cal T}_{\mathrm L}$    & $\mathcal{T}_{\mathrm{CL}}$ & ${\cal T}_{\mathrm {SL}}$ & ${\cal T}_{\mathrm {CSL}}$           \\ [1ex]
			\hline
			Case I & simple & 4.90 & 5.34 & 4.94 & 5.01 \\
			& permuted block  & 3.29 & 5.00 & 5.07 & 4.92 \\
			& minimization & 3.36 & 5.03 & 5.01 & 5.24 \\ \hline 
			Case II & simple & 5.05 & 5.59 & 5.16 & 5.28 \\
			& permuted block & 4.19 & 5.37 & 4.86 & 4.92 \\
			& minimization & 4.04 & 5.42 & 4.92 & 5.27 \\ \hline 
			Case III & simple & 4.88 & 5.53 & 5.12 & 5.18 \\
			& permuted block & 2.98 & 5.44 & 5.28 & 5.32 \\
			& minimization & 3.28 & 5.60 & 5.46 & 5.52 \\ \hline 
			Case IV & simple & 4.85 & 5.36 & 5.19 & 5.39 \\
			& permuted block & 4.16 & 5.19 & 4.87 & 4.96 \\
			& minimization & 4.64 & 5.64 & 5.35 & 5.38   \\	
			\hline 
		\end{tabular}
	\end{center}
\end{table}

\begin{figure}[h]
	\centering
	\includegraphics[scale=0.6]{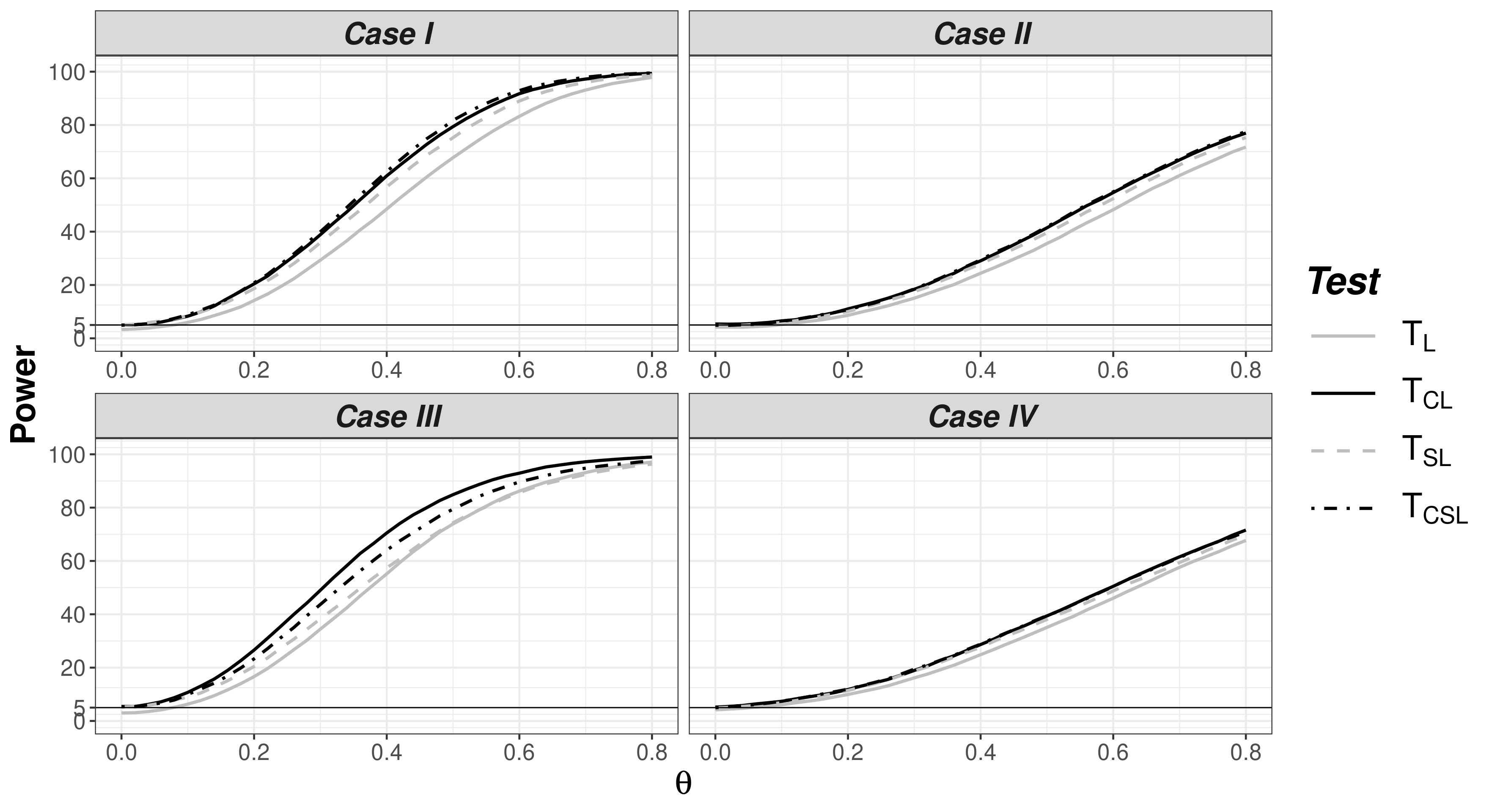}
\end{figure}
\begin{figure}[h]
	\includegraphics[scale=0.6]{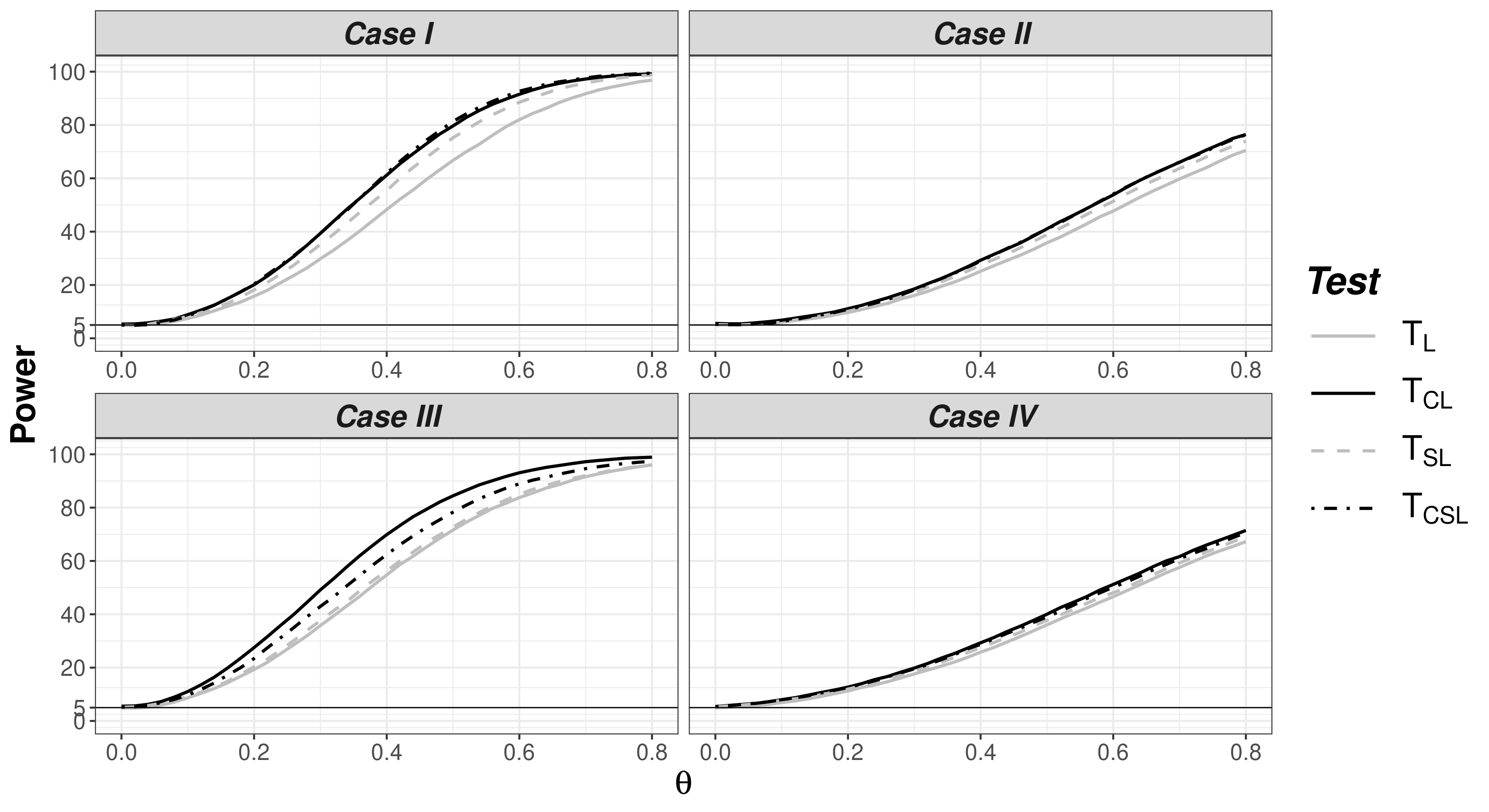}
	\includegraphics[scale=0.6]{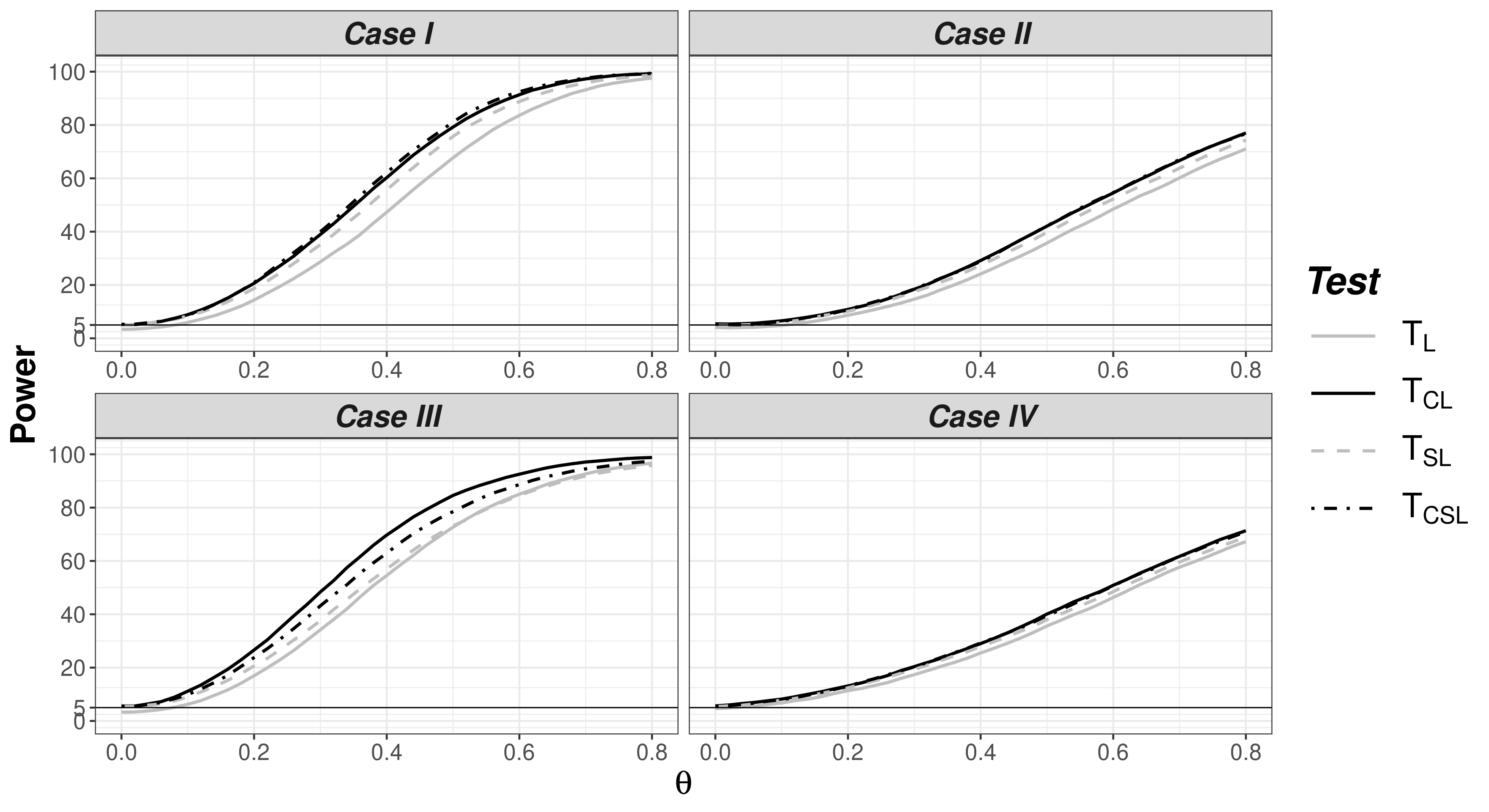}
	\caption{Power curves based on 10,000 simulations  with $n=200$ under permuted block randomization (previous page), simple randomization (top this page), and minimization (bottom this page).\label{fig: n=200}} 
\end{figure}

\subsection{Simulations under violations of Assumption CR}
This simulation setting is the same as Case III, except that $C_j$ follows a Cox model with conditional hazard $\log (1.1) \exp (-\psi j+ \eta_C^\top W)$ for $j=0,1$, $\psi$ ranges from 0 to 1 for different extent of assumption violation, and $\eta_C = (0.2, 0.2, 0.2)^\top$. Type I error rates for four tests under three randomization schemes are shown in Table \ref{tb: sim cr violated}. It can be seen that type I error is inflated as $\psi$   becomes larger.

\begin{table}[ht]
	\centering
	\caption{Type I errors (in \%) based on 10,000 simulations with $n=500$   when Assumption CR is violated.\label{tb: sim cr violated}}
	\begin{tabular}{lccccc}	\hline \\[-1.5ex]
		\multicolumn{1}{c}{Randomization}  		&    $\psi$   & ${\cal T}_{\mathrm L}$    & ${\cal T}_{\mathrm {CL}}$ & ${\cal T}_{\mathrm {SL}}$ & ${\cal T}_{\mathrm {CSL}}$           \\ [1ex]\hline
		simple         & 0.0 & 4.89 & 5.14  & 5.12 & 4.80 \\
		& 0.1 & 4.84 & 5.21  & 5.22 & 4.93 \\
		& 0.2 & 5.00 & 5.23  & 5.15 & 5.12 \\
		& 0.3 & 5.27 & 5.59  & 5.29 & 5.03 \\
		& 0.4 & 5.35 & 5.96  & 5.41 & 5.30 \\
		& 0.5 & 5.84 & 6.51  & 5.34 & 5.40 \\
		& 0.6 & 6.34 & 7.25  & 5.35 & 5.75 \\
		& 0.7 & 6.79 & 7.99  & 5.76 & 5.82 \\
		& 0.8 & 7.85 & 8.95  & 5.94 & 5.93 \\
		& 0.9 & 8.63 & 10.07 & 6.12 & 6.43 \\
		& 1.0 & 9.70 & 11.60 & 6.52 & 6.94 \\ \hline 
		permuted block & 0.0 & 3.41 & 5.36  & 5.40 & 4.86 \\
		& 0.1 & 3.38 & 5.49  & 5.43 & 5.04 \\
		& 0.2 & 3.37 & 5.36  & 5.36 & 4.97 \\
		& 0.3 & 3.63 & 5.79  & 5.41 & 5.20 \\
		& 0.4 & 3.87 & 6.04  & 5.44 & 5.20 \\
		& 0.5 & 4.27 & 6.76  & 5.60 & 5.30 \\
		& 0.6 & 4.62 & 7.08  & 5.69 & 5.43 \\
		& 0.7 & 5.28 & 7.68  & 5.84 & 5.68 \\
		& 0.8 & 5.78 & 8.84  & 6.06 & 5.95 \\
		& 0.9 & 6.53 & 9.90  & 6.35 & 6.39 \\
		& 1.0 & 7.52 & 11.34 & 6.71 & 6.81 \\ \hline 
		minimization   & 0.0 & 3.13 & 5.26  & 5.15 & 5.10 \\
		& 0.1 & 3.11 & 5.27  & 5.10 & 5.08 \\
		& 0.2 & 3.28 & 5.54  & 5.26 & 5.14 \\
		& 0.3 & 3.42 & 5.72  & 5.18 & 4.97 \\
		& 0.4 & 3.63 & 6.19  & 5.08 & 5.10 \\
		& 0.5 & 4.10 & 6.87  & 5.17 & 5.27 \\
		& 0.6 & 4.45 & 7.46  & 5.55 & 5.38 \\
		& 0.7 & 5.14 & 8.10  & 5.55 & 5.65 \\
		& 0.8 & 5.74 & 9.02  & 5.69 & 6.06 \\
		& 0.9 & 6.48 & 10.15 & 5.92 & 6.25 \\
		& 1.0 & 7.39 & 11.15 & 6.50 & 6.57\\\hline 
	\end{tabular}
\end{table}

\clearpage

\section{Lemmas and Additional theoretical results}

\subsection{Asymptotic optimality}

Consider a general class of log-rank score functions
\begin{align*}
	\hat U_{\mathrm{CL}} (b_0, b_1) &= \frac{1}{n} \sum_{i=1}^{n} \left[ I_i  \{ \olim_{i1} - (X_i- \bar X)^\top b_1 \} -  (1-I_i) \{ \olim_{i0}  - (X_i- \bar X)^\top b_0\} \right]  + n^{-1/2} o_p(1) ,
\end{align*}
where $b_1$ and $b_0$ are any fixed constants. The following theorem  derives the asymptotic distribution of $	\hat U_{\mathrm {CL}} (b_0, b_1)$, and shows that $	\hat U_{\mathrm {CL}} (\beta_0, \beta_1)$ has the smallest variance. 
\begin{theorem} \label{theo: b0 b1}
	Assume {(D)}, {(D$^\dag$)}, and that all levels of $Z$ used in covariate-adaptive randomization are included in $X_i$ as a sub-vector.
	Then the following results hold. 
	\begin{align*}
		&\sqrt{n}  \left(   \hat  U_{\mathrm {CL}}(b_0, b_1) -\frac{n_1\theta_1 -n_0\theta_0}{n}\right)   \xrightarrow{d}  N\left(0,\sigma_{\mathrm {CL}}^2 (b_0, b_1) \right),
	\end{align*}
	where \begin{align*}
		\sigma_{\mathrm {CL}}^2 (b_0, b_1) = & 	 \pi  E\left\{ \var ( \olim_{i1} - X_i^\top b_1  \mid Z_i)  \right\}  + (1- \pi ) E\left\{ \var ( \olim_{i0} - X_i^\top b_0  \mid Z_i)  \right\}  \\
		&\qquad +   \nu  \var \left\{E ( \olim_{i1} - X_i^\top b_1  \mid Z_i)  + E( \olim_{i0} - X_i^\top b_0  \mid Z_i)   \right\}    \\
		&\qquad +  \var \left\{ \pi E ( \olim_{i1} - X_i^\top b_1  \mid Z_i)  - (1-\pi) E( \olim_{i0} - X_i^\top b_0  \mid Z_i)   \right\}   \\
		&\qquad 	 +  \{ 2\pi \beta_1- 2(1-\pi)\beta_0 - \pi b_1 + (1-\pi) b_0    \}^\top \var (X_i  ) \{\pi b_1 - (1-\pi)b_0\}   .
	\end{align*}
	Furthermore, 
	\begin{align*}
		&\sigma^2_{\mathrm {CL}} (b_0 ,b_1)  -\sigma^2_{\mathrm {CL}} (\beta_0 ,\beta_1)  \\
		&= \pi(1-\pi)  (\beta_1- b_1+ \beta_0 - b_0)^\top \left[ E\{\var (X_i \mid Z_i )\}   + \frac{\nu}{\pi(1-\pi)} \var \{ E(X_i \mid Z_i)\} \right](\beta_1- b_1+ \beta_0 - b_0) 
	\end{align*}
	which is  $\geq 0$ with strict inequality holding unless $\beta_1+\beta_0 = b_0+b_1$ or  $\var(X_i \mid Z_i )= 0$ and $ \nu=0$ almost surely.  
\end{theorem}

To end this section we emphasize that the use of  $\hat{O}_{ij}$'s  in (\ref{hatoij}) as
derived outcomes to reduce variability of $\hat U_{\mathrm L}$ is a key to our results.
The use of other derived outcomes may not achieve guaranteed efficiency gain. For example, \cite{tangen1999} and \cite{jiang2008} considered log-rank scores 
$ \tilde O_{ij}    = \int_{0}^\tau {\textstyle \frac{1}{2}} \big\{  dN_{ij} (t) -    Y_{ij} (t)  d\bar  N(t) / \bar Y(t)  \big\} $ as derived outcomes;
however, using $\tilde{O}_{ij}$ to replace $\hat{O}_{ij}$ in (\ref{CL1}) and (\ref{hatbeta}) produces an adjusted score that is not necessarily more efficient than  the unadjusted $\hat U_{\mathrm L}$ and is always less efficient than $\hat U_{\mathrm {CL}}$ in (\ref{CL1}) (as shown in Theorem S1),
due to the reason that using $\tilde{O}_{ij}$ instead of $\hat O_{ij}$ in (\ref{hatbeta})
does not correctly capture the true correlation between $\olim_{ij} $ and $X_i$. 
Furthermore, using $\tilde{O}_{ij}$ may not produce a valid test  under covariate-adaptive randomization,  
even with (C)-(D)  and all joint levels of $Z_i$ included in $X_i$.

\subsection{Covariate adjustment in hazard ratio estimation under Cox model}

After testing the null hypothesis of no treatment effect using the  covariate-adjusted  log-rank test $\mathcal{T}_{\mathrm{CL}}$, it is of interest to also report an effect size estimate and confidence interval. One common parameter is the hazard ratio $e^{\theta}$ under the Cox proportional hazards model 
\[
\lambda_1(t)  = \lambda_0 (t) e^{\theta}. 
\]
Without using covariates, the score equation from the partial likelihood is 
\[
\widehat U_{\mathrm {L}} (\vartheta) = \frac1n  \sum_{i=1}^n \int_0^\tau \left\{ I _i - \frac{S^{(1)} (\vartheta, t)  }{S^{(0)} (\vartheta, t) }\right\} dN_i(t) ,
\]
where $S^{(1)} (\vartheta, t) = n^{-1} \sum_{i=1}^n I_i Y_i(t) e^{\vartheta I_i } = e^{\vartheta } \bar Y_1(t) $ and  $S^{(0)} (\vartheta, t) = n^{-1} \sum_{i=1}^n Y_i(t) e^{\vartheta I_i }  = e^{\vartheta } \bar Y_1(t)  + \bar Y_0(t) $. The log-rank test uses $\widehat U_{\mathrm {L}} (0) $.

Let 
\begin{align*}
	O_{i1} (\theta) & = \int_0^\tau \left\{ 1- \frac{s^{(1)} (\theta, t)}{s^{(0)} (\theta, t)}\right\} \left\{ dN_{i1}(t) - Y_{i1}(t) e^{\theta} \frac{E(dN_{i}(t))}{s^{(0)} (\theta, t)}\right\} \\
	O_{i0} (\theta) & = \int_0^\tau  \frac{s^{(1)} (\theta, t)}{s^{(0)} (\theta, t)} \left\{ dN_{i0}(t) - Y_{i0}(t) \frac{E(dN_{i}(t))}{s^{(0)} (\theta, t)}\right\} 	,
\end{align*}
where $s^{(1)}(\theta , t) = e^{\theta} \mu(t) E\{  Y_i(t)\} $ and $s^{(0)} (\theta, t) = \{ e^{\theta} \mu(t) +1- \mu(t)\} E\{  Y_i(t)\}$. 

\begin{theorem} \label{theo: hazard estimation Cox model}
	Assume (C) and (D), and all joint levels of $Z_i$ used in covariate-adaptive randomization are included in $X_i$ as a sub-vector. Also assume the 
	Cox proportional hazards model $\lambda_1(t)  = \lambda_0 (t) e^{\theta}$. Then, the following results hold regardless of which covariate-adaptive randomization scheme is applied.
	$$
	\sqrt{n} (\widehat\theta_{\mathrm {CL}}  - \theta ) \xrightarrow{d} N \left ( 0,\frac{\sigma_{\mathrm {CL}}^2 (\theta )}{\sigma^4_{\mathrm {L}} (\theta )} \right) ,
	$$
	where 		$\sigma_{\mathrm {CL}}^2(\theta )   = \sigma_{\mathrm {L}}^2 (\theta)  - \pi(1-\pi) (\beta_1 (\theta)+\beta_0 (\theta) )^\top \Sigma_X   (\beta_1 (\theta)+\beta_0 (\theta) )$, 
	\begin{align*}
		\sigma_{\mathrm {CL}}^2 (\theta) & = \pi \mathrm{var} (O_{i1} (\theta) - X_i^\top \beta_1 (\theta) ) +  (1-\pi ) \mathrm{var} (O_{i0} (\theta) - X_i^\top \beta_0 (\theta) )  \\
		&\quad + (\pi\beta_1(\theta)  - (1-\pi) \beta_0(\theta ) )^\top \Sigma_X (\pi\beta_1(\theta)  - (1-\pi) \beta_0(\theta ) ) , \\
		\sigma_{\mathrm {L}}^2 (\theta ) & = \pi \var (O_{i1} (\theta)) + (1-\pi) \var (O_{i0}(\theta)), 
	\end{align*}
	and  $\beta_j (\theta) =  \Sigma_X^{-1} \mathrm{cov} (X_i , O_{ij} (\theta))$ for $j=0,1$.
\end{theorem}

Inference can be made based on Theorem \ref{theo: hazard estimation Cox model} and estimated variance  $\hat\sigma^2_{\mathrm{CL}} (\hat\theta_{\mathrm {CL}})/ \hat\sigma^4_{\mathrm{L}} (\hat\theta_{\mathrm {CL}}) $, with 
$$
\hat\sigma^2_{\mathrm{L}} (\hat\theta_{\mathrm {CL}})= - \partial \widehat U_{\mathrm L} (\vartheta ) / \partial \vartheta \mid_{\vartheta =\hat\theta_{\mathrm {CL}} } =  
n^{-1} \sum_{i=1}^n \int_0^\tau \frac{e^{\hat\theta_{\mathrm {CL}} } \bar Y_1(t) \bar Y_0(t)}{\{ e^{\hat\theta_{\mathrm {CL}} } \bar Y_1(t) +\bar Y_0(t)\}^2} dN_i(t)
$$
and  $\hat\sigma^2_{\mathrm{CL}} (\hat\theta_{\mathrm {CL}})= \hat\sigma^2_{\mathrm{L}} (\hat\theta_{\mathrm {CL}}) - \pi(1-\pi) (\hat \beta_1(\hat\theta_{\mathrm {L}}) )+ \hat\beta_0(\hat\theta_{\mathrm {L}}) )^\top \hat \Sigma_X (\hat\beta_1(\hat\theta_{\mathrm {L}}) )+\hat\beta_0(\hat\theta_{\mathrm {L}}) ) $.

\subsection{Covariate adjustment in hazard ratio estimation under stratified Cox model}

In this section, we assume the stratified Cox proportional hazards model:
\[
\lambda_{z1}(t) = \lambda_{z0} (t) e^\theta. 
\]

Without using covariates, the score equation from the partial likelihood is 
\[
\widehat U_{\mathrm{SL}} (\vartheta) = n^{-1} \sum_{z} \sum_{i: Z_i =z}  \int_0^\tau \bigg\{ I _i - \frac{S_z^{(1)} (\vartheta, t)  }{S_z^{(0)} (\vartheta, t) }\bigg\} dN_i(t) .
\]
where $S_z^{(1)} (\vartheta, t) = n^{-1} \sum_{i: Z_i = z} I_i Y_i(t)e^{\vartheta I_i } =  e^{\vartheta } \bar Y_{z1}(t) $ and  $S_z^{(0)} (\vartheta, t) = n^{-1} \sum_{i: Z_i = z} Y_i(t) e^{\vartheta I_i }   = e^{\vartheta } \bar Y_{z1}(t)  + \bar Y_{z0}(t)  $. The log-rank test uses $\widehat U_{\mathrm {SL}} (0) $. The maximum partial likelihood estimator $\widehat\theta_{\mathrm {SL}} $ of $\theta$ is 
a solution to $\widehat U_{\mathrm {SL}} ( \vartheta ) = 0$. Our covariate-adjusted score  is 
\begin{align*}
	\widehat U_{\mathrm {CSL}} (\vartheta ) & = \widehat U_{\mathrm {SL}} (\vartheta ) -  \frac{1}{n} \sum_z \sum_{i: Z_i= z}   \left\{ I_i  (X_i- \bar X_z)^\top \hat\gamma_1 (\hat \theta_{\mathrm {SL}}) - (1-I_i)    ( X_i- \bar X_z)^\top \hat \gamma_0  (\hat \theta_{\mathrm {SL}})   \right\} , 
\end{align*}
where, for $j=0,1$, $		\widehat \gamma_j (\widehat \theta_{\mathrm {SL}}) $ is equal to 
$	\widehat \gamma_j $ in the main article with $ \widehat O_{zij}$ replaced by   
$$ 
\widehat O_{zij}(\widehat \theta_{\mathrm {SL}})  = \int_0^\tau  \frac{\{ e^{\widehat \theta_{\mathrm {SL}}} \bar Y_{z1}(t) \}^{(1-j)} \{ \bar Y_{z0}(t)\}^{j}  }{e^{\widehat \theta_{\mathrm {SL}}}  \bar Y_{z1}(t) + \bar Y_{z0} (t)}  \left\{ dN_{ij}(t) - \frac{  Y_{ij}(t) e^{j \widehat \theta_{\mathrm {SL}}}  d \bar N_z(t)}{e^{\widehat \theta_{\mathrm {SL}}}  \bar Y_{z1}(t) + \bar Y_{z0} (t) }\right\} .
$$   
We obtain $\hat \theta_{\mathrm{CSL}}$ from solving $	\widehat U_{\mathrm {CSL}} (\vartheta )  = 0$.

Let 
\begin{align*}
	O_{zi1} (\theta) & = \int_0^\tau \left\{ 1- \frac{s_z^{(1)} (\theta, t)}{s_z^{(0)} (\theta, t)}\right\} \left\{ dN_{i1}(t) - Y_{i1}(t)e^\theta  \frac{E(dN_{i}(t) \mid Z_i=z)}{s_z^{(0)} (\theta, t)}\right\} \\
	O_{zi0} (\theta) & = \int_0^\tau  \frac{s_z^{(1)} (\theta, t)}{s_z^{(0)} (\theta, t)} \left\{ dN_{i0}(t) - Y_{i0}(t) \frac{E(dN_{i}(t)\mid Z_i = z)}{s_z^{(0)} (\theta, t)}\right\} 	,
\end{align*}
where $s_z^{(1)}(\theta , t) =  e^{\theta}  \mu_z(t) E\{  Y_i(t)\} $ and $s_z^{(0)} (\theta, t) = \{  e^{\theta}  \mu_z(t) +1- \mu_z(t)\} E\{  Y_i(t)\}$.

\begin{theorem} \label{theo: hazard estimation stratified Cox model}
	Under (C-z) and (D). Also assume the Cox proportional hazards model $\lambda_{z1}(t)  = \lambda_{z0} (t) e^{\theta}$. Then, the following results hold regardless of which covariate-adaptive randomization scheme is applied.
	$$
	\sqrt{n} (\widehat\theta_{\mathrm {CSL}}  - \theta ) \xrightarrow{d} N \left ( 0,\frac{\sigma_{\mathrm {CSL}}^2 (\theta )}{\sigma^4_{\mathrm {SL}} (\theta )} \right) ,
	$$
	where 		$\sigma_{\mathrm {CSL}}^2(\theta )   = \sigma_{\mathrm {SL}}^2 (\theta)  - \pi(1-\pi) (\gamma_1 (\theta)+\gamma_0 (\theta) )^\top E\{ \var (X_i\mid Z_i)\}  (\gamma_1 (\theta)+\gamma_0 (\theta) )$, $	\sigma_{\mathrm {SL}}^2 (\theta )  = \sum_{z} \mathrm{pr} (Z_i = z) \{ \pi \var (O_{zi1} (\theta)\mid Z_i = z) + (1-\pi) \var (O_{zi0}(\theta)\mid Z_i = z) \} $.
\end{theorem}

The proof of Theorem \ref{theo: hazard estimation stratified Cox model} is similar to the proof of Theorem \ref{theo: hazard estimation Cox model} and thus is omitted. Based on Theorem \ref{theo: hazard estimation stratified Cox model},  inference can be made based on the estimated variance  $\hat\sigma^2_{\mathrm{CSL}} (\hat\theta_{\mathrm {CSL}})/ \hat\sigma^4_{\mathrm{SL}} (\hat\theta_{\mathrm {CSL}}) $, with
\begin{align*}
	\hat\sigma^2_{\mathrm{SL}} (\hat\theta_{\mathrm {CSL}}) & = - \partial \widehat U_{\mathrm {SL}} (\vartheta ) / \partial \vartheta \mid_{\vartheta =\hat\theta_{\mathrm {CSL}} } =  
	n^{-1} \sum_{z}\sum_{i: Z_i =z} \int_0^\tau \frac{e^{\hat\theta_{\mathrm {CSL}} } \bar Y_{z1}(t) \bar Y_{z0}(t)}{\{ e^{\hat\theta_{\mathrm {CSL}} } \bar Y_{z1}(t) +\bar Y_{z0}(t)\}^2} dN_i(t)
\end{align*}
and  $\hat\sigma^2_{\mathrm{CSL}} (\hat\theta_{\mathrm {CSL}})= \hat\sigma^2_{\mathrm{SL}} (\hat\theta_{\mathrm {CSL}}) - \pi(1-\pi) (\hat \gamma_1(\hat\theta_{\mathrm {SL}}) + \hat\gamma_0(\hat\theta_{\mathrm {SL}}) )^\top \{ \sum_{z} (n_z/n) \hat \Sigma_{X\mid z} \} (\hat\gamma_1(\hat\theta_{\mathrm {SL}}) +\hat\gamma_0(\hat\theta_{\mathrm {SL}}) ) $.

\subsection{Lemmas}
The following lemmas are useful for proving the main theorems. 

\begin{lemma} \label{lemma: beta}
	Under condition (D),\\
	(a)  $ \hat\beta_1   = \beta_1  + o_p(1) $ and  $ \hat\beta_0  = \beta_0  + o_p(1) $. \\
	(b) $ \hat\gamma_1  = \gamma_1  + o_p(1) $ and  $ \hat\gamma_0  = \gamma_0  + o_p(1) $. 
\end{lemma}


\begin{lemma} \label{lemma: S3}
	Assume (CR) and (D). 	Let $E_{H_0^\dagger}$, $\lambda_{H_0^\dagger}$, $\mu_{H_0^\dagger}(t)$, and $p_{H_0^\dagger}(t)$ be 
	the expectation, hazard, $\mu(t)$, and $p(t)$ under the null hypothesis $H_0^\dagger: \lambda_1(t, v) = \lambda_0(t, v)$ for all $t$ and $v$, where $\lambda_j(t, v)$ is the true hazard function of $T_j$ conditional on  $V=v$ for $j=0,1$, and $V$ is in (CR).  Then, for any $t$, \\
	(a) $	E_{H_0^\dagger}\{ I_i Y_i (t) \mid V_i  \} = \pi E_{H_0^\dagger}\{ Y_{i1} (t)  \mid V_i  \}  = \mu_{H_0^\dagger}(t) E_{H_0^\dagger}\{ Y_i(t)  \mid V_i  \} , $\\
	(b) $E_{H_0^\dagger}\{(1- I_i) Y_i (t) \mid V_i  \} = (1-\pi) E_{H_0^\dagger}\{ Y_{i0} (t)  \mid V_i  \}  =\{1-\mu_{H_0^\dagger}(t)\}  E_{H_0^\dagger}\{ Y_i(t)  \mid V_i  \} , $  \\
	(c) $ E_{H_0^\dagger} \{ Y_{ij} (t) \lambda_{H_0^\dagger} (t, V_i)\} =  p_{H_0^\dagger}(t) E_{H_0^\dagger} \{ Y_{ij}(t) \}, \  j=0,1. $
\end{lemma}
Note that conditions (C) and (D) imply Lemma S2 to hold with $V=\emptyset$. In addition, as (TR) leads to the equivalence of $H_0$ and $H_0^\dagger$, thus, (CR), (TR) and (D) imply Lemmas \ref{lemma: S3}(a)-(c) to also hold under $H_0$.

\begin{lemma}  \label{lemma: S0}
	Let $\tilde \sigma_{\mathrm {L}}^2  = \int_0^\tau\mu(t) \{1-\mu(t)\} E\{ dN_i(t)\} $. 
	Assume (CR) and (D),	$\hat\sigma_{\mathrm {L}}^2 \xrightarrow{p} \tilde \sigma_{\mathrm {L}}^2 $ and  
	$$
	\tilde \sigma_{\mathrm {L}}^2  	= \int_0^\tau E\left[  \mu(t) \{1-\mu(t)\} \{\pi Y_{i1}(t)\lambda_1(t, V_i) + (1- \pi) Y_{i0}(t)\lambda_0(t, V_i)\}  \right] dt . 
	$$
\end{lemma}

\begin{lemma}  \label{lemma: S1}
	Assume the conditions in Theorem 1 and the local alternative hypothesis specified in Theorem 1(c). 
	Then, 
	$E(O_{ij} ) \to 0 $, $j=0,1$, and 	both $\sigma^2_{\mathrm {L}}$ and 
	$
	\tilde \sigma_{\mathrm {L}}^2 \to  \pi \var_{H_0}(O_{i1}) + (1-\pi )\var_{H_0}(O_{i0}) $,
	where $\var_{H_0}$ denotes the  variance under $H_0$.
\end{lemma}

%
%

%
%

\section{Technical Proofs}

\subsection{Proofs of Lemmas}

\noindent
{\sc Proof of Lemma \ref{lemma: beta}}. \\
(a)  
We show the proof for $\hat\beta_1$ and $\beta_1$. Note first that
\begin{align*}
	\frac{1}{n_1} \sum_{i=1}^{n} I_i (X_i - \bar X_1) (X_i - \bar X_1)^\top \xrightarrow{p} \Sigma_X 
\end{align*}
from the proof of Lemma 3 in \cite{ye2021better}. 	From Lemma 3 of \cite{Ye:2020survival}, we have that $\bar Y_0(t) \xrightarrow{p} (1-\pi) E\{ Y_{i0}(t)\},  \bar Y_1(t) \xrightarrow{p} \pi E\{ Y_{i1}(t)\}, \bar Y(t) \xrightarrow{p}  E\{Y_i(t)\}$. Similarly, we can show that $n_1^{-1} \sum_{i=1}^{n} I_i X_i dN_{i1}(t) \xrightarrow{p} E\{X_i dN_{i1}(t)\} $, $n_1^{-1} \sum_{i=1}^{n} I_i X_i Y_{i1}(t) \xrightarrow{p} E\{X_i Y_{i1}(t)\} $, and $\bar N(t) \xrightarrow{p} E\{ N_i(t)\}$.  Hence,
\begin{align*}
	\frac{1}{n_1}	\sum_{i=1}^{n} I_i (X_i - \bar X_1) \hat O_{i1} \xrightarrow{p} \cov (X_i, \olim_{i1} ) ,
\end{align*}
concluding the proof that  $ \hat\beta_1  = \beta_1  + o_p(1) $. The result for  $\hat\beta_0$ can be shown in the same way. \\	
(b) The proof for $\hat\gamma_1,  \hat\gamma_0$ and $\gamma_1, \gamma_0$ are similar and can be established from showing
\begin{align*}
	\frac{1}{n_1}  \sum_z \sum_{i: Z_i=z} I_i (X_i -  \bar X_{z1}  ) (X_i -  \bar X_{z1}  )^\top &  \xrightarrow{p} E\{\var (X_i \mid Z_i )\} \\
	\frac{1}{n_1}  \sum_z \sum_{i: Z_i=z} I_i (X_i -  \bar X_{z1}  ) \hat O_{zi1}   & \xrightarrow{p}  \sum_{z}P(Z=z)   \cov\left(X_i,   O_{zi1}  \mid Z_i=z \right)  \\
	&= \cov\left(X_i, \sum_{z} I(Z_i=z) ( O_{zi1} - \theta_{z1} ) \right), 
\end{align*}
where $ \gamma_j = E\{\var (X_i \mid Z_i )\}^{-1}   \cov\left(X_i, \sum_{z} I(Z_i=z) ( O_{zij} - \theta_{zj} ) \right)$. \\

\noindent
{\sc Proof of Lemma \ref{lemma: S3}}. \\
For simplicity, we remove the subscript $H_0^\dagger$ and assume all calculations are under $H_0^\dagger$. \\
(a) Note that 
\begin{align*}
	&\pi E\{ Y_{i1} (t)  \mid V_i \} = E  \left[ E(I_i \mid Z_1,\dots, Z_n, V_i  ) E\{ Y_{i1} (t) \mid Z_1,\dots, Z_n, V_i\}  \mid V_i \right]  \\
	&= E\left[  I_i Y_{i} (t)  \mid V_i  \right] = E\left[ E\{ I_i \mid Y_i(t) =1, V_i\}Y_{i} (t) \mid V_i \right] \\
	&= \mu(t) E\left\{ Y_{i} (t) \mid V_i \right\},
\end{align*} 
where the first equality is because of $E(I_i\mid Z_1,\dots, Z_n, V_i)= E(I_i\mid Z_1,\dots, Z_n) = \pi$, the second equality is because of the conditional independence $I_i \perp Y_{i1}(t) \mid Z_1,\dots, Z_n, V_i $ implied by (D), the last equality is from $E\{ I_i \mid Y_i(t) =1, V_i\} = \mu(t)$ under (CR) and $H_0^\dagger$ due to Lemma 1 in \cite{Ye:2020survival}. \\
(b) The proof is the same as that for (a). \\
(c) The result is straightforward from  showing that 
\begin{align*}
	E\{ Y_{i1}(t) \lambda(t, V_i)\}  & =  \pi^{-1} E\{ I_i Y_{i}(t) \lambda(t, V_i)\}= \pi^{-1} E\left[ E\{ I_i \mid Y_i(t) = 1, V_i\} Y_i(t) \lambda(t, V_i)\right] \\
	&= \pi^{-1} \mu(t) E\left[ Y_i(t) \lambda(t, V_i)\right] =  \pi^{-1} \mu(t) E\left[ Y_i(t)\right] p(t)\\
	E\{ Y_{i1}(t) \}  & =  \pi^{-1} E\{ I_i Y_{i}(t) \}= \pi^{-1} E\left[ E\{ I_i \mid Y_i(t) = 1\} Y_i(t) \right] \\
	&= \pi^{-1} \mu(t) E\left[ Y_i(t)\right],
\end{align*}
as well as the counterparts for $E\{ Y_{i0}(t) \lambda(t, V_i)\}  $ and $E\{ Y_{i0}(t) \}  $. 
\\

\noindent
{\sc Proof of Lemma \ref{lemma: S0}}. \\
The first result	$\hat\sigma_{\mathrm {L}}^2 \xrightarrow{p} \tilde \sigma_{\mathrm {L}}^2  $ is because $\bar Y_1(t) \bar Y_0(t)/ \bar Y(t)^2 \xrightarrow{p} \mu(t) \{ 1- \mu(t)\} $ from Lemma 3 of \cite{Ye:2020survival}, 
and $d \bar N(t) \xrightarrow{p} E\{ d N_i(t)\}$. 

Then, under (CR), from the theory of counting processes in survival analysis \citep{Andersen:1982aa}, 
the process $N_{ij}(t)$ has random intensity process of the form $Y_{ij} (t) \lambda_j(t, V_i)$, $j=0,1$.  Hence, for $i=1,..., n$, $j=0,1,$
the process $N_{ij} (t)  - \int_{0}^{t} Y_{ij}(s) \lambda_j(s, V_i) ds $  is a local square integrable martingale  with respect to the filtration ${\cal F}_t  = \sigma\{ N_{ij} (u), (1- \delta_{ij} ) {\cal I}(X_{ij}\leq u), V_i: 0\leq u\leq t \}$. From the fact that martingales have expectation zero, we conclude that $	E\{ dN_{ij} (t) \}  =E \{  Y_{ij} (t) \lambda_j(t, V_i)\} dt$. The second result follows from
\begin{align*}
	E\{ dN_{i} (t) \} &=E\{ I_i dN_{i1} (t)+ (1-I_i) dN_{i0}(t)\}  \\
	&= \pi E \{  Y_{i1} (t) \lambda_1(t, V_i)\} dt +(1-\pi) E \{  Y_{i0} (t) \lambda_0(t, V_i)\} dt. 
\end{align*}

\noindent
{\sc Proof of Lemma \ref{lemma: S1}}. \\
Let  $V_i=\emptyset$ under (C), and $V_i $ is the $V_i$ in (CR) under (CR)-(TR). Then,
$$			E(O_{i1}) =  \int_0^\tau E[  \{1-\mu(t) \}  Y_{i1}(t) \{  \lambda_1(t, V_i) - p(t)\}]  dt . $$
Under the local alternative and from the dominated convergence theorem, for every $t$,
\begin{align}
	E\{ Y_{i1}(t)  \} &= \int e^{- \int_0^t \lambda_1(s, v) ds } \P (C_1\geq t\mid V=v)  dF(v)\nonumber\\
	&\rightarrow   \int  e^{- \int_0^t \lambda_{H_0^\dagger}(s, v) ds } \P (C_1\geq t\mid V=v) dF(v) \nonumber \\
	&=  E_{H_0^\dagger}\{ Y_{i1}(t)  \}, \label{eq: 1}
\end{align}
where $F$ is the distribution of  $V$ and 
$\lambda_{H_0^\dagger}$ and  $E_{H_0^\dagger}$ denote the hazard and expectation under $H_0^\dagger$ defined in Lemma \ref{lemma: S3}, respectively. Similarly, we can show that 
\begin{align}
	&E\left\{ Y_{i1}(t) \lambda_1(t, V_i)  \right\} \rightarrow E_{H_0^\dagger} \left\{ Y_{i1}(t) \lambda_{H_0^\dagger}(t, V_i)   \right\} .  \label{eq: 2}
\end{align}
These imply that $\mu(t) \rightarrow \mu_{H_0^\dagger}(t)$ and $p(t) \rightarrow p_{H_0^\dagger}(t)$, where $\mu_{ H_0^\dagger} (t) $ and $p_{H_0^\dagger}(t)$ are $\mu(t)$ and  $p(t)$ under $H_0^\dagger$, respectively.   Hence, again from the dominant convergence theorem,
\begin{align*}
	E(O_{i1}) &= \int_0^\tau \{1-\mu(t) \} E[   Y_{i1}(t) \{  \lambda_1(t, V_i) - p(t)\}]  dt \\
	&\to  \int_0^\tau \{1-\mu_{H_0^\dagger }(t)\}  E_{H_0^\dagger }[   Y_{i1}(t) \{  \lambda_{H_0^\dagger }(t, V_i) - p_{H_0^\dagger }(t)\}]  dt \\
	& = E_{H_0^\dagger } (O_{i1}) \\
	&= 0
\end{align*}
where the last equality follows from Lemma \ref{lemma: S3}(c). The proof for 
$E(O_{i0} ) \to 0 $ is the same. 

Let $\var_{H_0}$ denote the variance under $H_0$. Theorem \ref{theo: b0 b1} with $b_0=b_1=0$ implies that $\sigma_{\mathrm {L}}^2 = \pi \var (O_{i1}) + (1-\pi) \var (O_{i0})  $.  Next, we show that under the local alternative, $\sigma_{\mathrm {L}}^2 \to \pi \var_{H_0^\dagger} (O_{i1}) + (1-\pi) \var_{H_0^\dagger} (O_{i0})$. 
For $\var (O_{ij}) $, since  $\var (O_{ij}) = E(O_{ij}^2) - \{ E(O_{ij})\}^2 $ and $E(O_{ij}) \to 0$,  it suffices to show that $E(O_{ij}^2)\to \var_{H_0^\dagger}(O_{ij}^2)$. 
Note that 
\begin{align*}
	E(O_{i1}^2 )  & = E\left[ \bigg\{ \int_0^\tau \{ 1- \mu(t)\} dN_{i1}(t)\bigg\}^2 \right. \\
	& \quad - 2 \int_0^\tau \int_0^\tau  \{ 1- \mu(t)\} Y_{i1} (t)  \lambda_1(t, V_i) \{ 1- \mu(s)\} Y_{i1} (s) p(s) dt ds  \\
	&
	\left. \quad +  \int_0^\tau \int_0^\tau  \{ 1- \mu(t)\} \{ 1- \mu(s)\} Y_{i1} (s) Y_{i1} (t)  p(s)  p(t) dt ds
	\right]  \\
	&=   \int_0^\tau \{ 1- \mu(t)\}^2 E\left\{ dN_{i1}(t) \right\}  \\
	&\qquad  - 2 \int_0^\tau \int_{t\geq s}  \{ 1- \mu(t)\} \{ 1- \mu(s)\}   p(s) E \{  Y_{i1} (t)  \lambda_1(t, V_i)  - Y_{i1} (t) p(t) \} 	dt ds  	,
\end{align*}
where the second equality is  because, when $t> s$, $Y_{i1} (t) d N_{i1} (s) = 0$,  $E\{ Y_{i1} (s) dN_{i1} (t)\mid \mathcal{F}_{t-}\} = Y_{i1} (s)  Y_{i1}(t) \lambda_1 (t, V_i) dt  $ for $t\geq s$, and $Y_{i1}(t) Y_{i1} (s) = Y_{i1} \{\max (t, s)\}$. These techniques will be used frequently in the following proofs, and will not be further elaborated. From  \eqref{eq: 1}-\eqref{eq: 2}, we have that $E \{  Y_{i1} (t)  \lambda_1(t, V_i)  - Y_{i1} (t) p(t) \} \to 0 $ for every $t $, and consequently $$E(O_{i1}^2 ) \to \int_0^\tau \{ 1- \mu_{H_0^\dagger}(t)\}^2 E_{H_0^\dagger}\left\{ Y_{i1} (t)   \lambda_{H_0^\dagger}(t, V_i) \right\}  dt ,$$
which is equal to $\var_{H_0^\dagger}(O_{i1})$ by the same argument and the fact that 
$E_{H_0^\dagger} (O_{i1}) = 0 $. 
Similarly, we can show that $E(O_{i0}^2 ) \to \int_0^\tau  \mu_{H_0^\dagger}(t)^2 E_{H_0^\dagger}\left\{ Y_{i0} (t)   \lambda_{H_0^\dagger}(t, V_i) \right\}  dt  = \var_{H_0^\dagger}(O_{i0})$. This concludes the proof that $\sigma_{\mathrm {L}}^2 \to \pi \var_{H_0^\dagger} (O_{i1}) + (1-\pi) \var_{H_0^\dagger} (O_{i0})  $ under the local alternative. 

For $\tilde \sigma_{\mathrm {L}}^2$,  under the local alternative,  from Lemma \ref{lemma: S3}, \eqref{eq: 1}-\eqref{eq: 2}, and a similar argument as above,
\begin{align*}
	\tilde \sigma_{\mathrm {L}}^2 &= \int_0^\tau \mu(t) \{ 1- \mu(t)\} \left[ \pi E\{ Y_{i1}(t)\lambda_1(t, V_i)\} + (1- \pi)E\{ Y_{i0}(t)\lambda_0(t, V_i)\}  \right] dt \\
	&  \to \int_0^\tau \mu_{H_0^\dagger}(t) \{ 1- \mu_{H_0^\dagger}(t)\} \left[ \pi E_{H_0^\dagger} \{ Y_{i1}(t)\lambda_{H_0^\dagger}(t, V_i)\} + (1- \pi)E_{H_0^\dagger}\{ Y_{i0}(t)\lambda_{H_0^\dagger}(t, V_i)\}  \right] dt  \\
	&=   \pi \var_{H_0^\dagger} (O_{i1}) + (1-\pi) \var_{H_0^\dagger} (O_{i0})    .
\end{align*}

The result follows from the fact that $H_0^\dagger = H_0$ under either (C) or (CR)-(TR).

\subsection{Proofs of Theorems}

\noindent
{\sc Proof of Theorem 1}. 

\noindent
(a) Following a Taylor expansion as in the Appendix of \cite{Lin:1989aa}, we obtain that under either the null or alternative hypothesis, 
\begin{align*}
	\hat U_{\mathrm {L}}  &= n^{-1} \sum_{i=1}^n \int_0^\tau \{ I_i -\mu(t)\} \{ dN_i(t) - Y_{i} (t) p(t) dt \}  + o_p(n^{-1/2})\\
	& = n^{-1} \sum_{i=1}^n \{ I_i \olim_{i1}  - (1- I_i) \olim_{i0} \} +o_p(n^{-1/2}),
\end{align*}
where $(\olim_{i1},\olim_{i0}) , i=1, \dots, n$ are i.i.d.. Then from  $   n^{-1} \sum_{i=1}^{n} I_i (X_i - \bar X) = {n^{-1} n_1(\bar X_1 - \bar  X)= O_p(n^{-1/2}) } $, and Lemma \ref{lemma: beta}, we have that 
\begin{align*}
	\hat U_{\mathrm {CL}}  =   n^{-1} \sum_{i=1}^n \left[   I_i \{ \olim_{i1}  - (X_i - \bar X)^\top \beta_1\}- (1- I_i)\{ \olim_{i0}  - (X_i - \bar X)^\top \beta_0\} \right] +o_p(n^{-1/2}).
\end{align*}
The rest of the proof is similar to the proof of Theorem 2 in \cite{ye2021better}. Define $ \mathcal{I} = \{I_1,\dots, I_n\} $ and $ \mathcal{S} = \{Z_1,\dots, Z_n\} $,  then 
\begin{align}
	&\hat U_{{\mathrm {CL}}} -  \left(\frac{n_1}{n} \theta_1 -\frac{n_0}{n} \theta_0 \right) \nonumber\\
	&= \frac1n \sum_{i=1}^{n} I_i \{ \olim_{i1}  -\theta_1 -  (X_i - \bar X)^\top \beta_1  \}- (1-I_i)  \{\olim_{i0} -\theta_0  -  (X_i -\bar X)^\top \beta_0\}  + o_p(n^{-1/2}) \nonumber \\
	&=  \frac1n \sum_{i=1}^{n} I_i \{ \olim_{i1} -\theta_1- (X_i - \mu_X)^\top \beta_1  \}- (1-I_i)  \{\olim_{i0} -\theta_0 -  (X_i - \mu_X)^\top \beta_0\}  \nonumber \\
	&\qquad +  \frac{n_1}{n} (\bar X- \mu_X)^\top \beta_1 - \frac{n_0}{n} (\bar X- \mu_X)^\top \beta_0 +o_p(n^{-1/2})\nonumber\\
	&=  \frac1n \sum_{i=1}^{n} I_i \{ \olim_{i1} - \theta_1 - (X_i - \mu_X)^\top \beta_1  \}-  \frac1n \sum_{i=1}^{n}(1-I_i)  \{\olim_{i0} - \theta_0-  (X_i - \mu_X)^\top \beta_0\}   \nonumber \\
	&\qquad +  \pi (\bar X- \mu_X)^\top \beta_1 - (1-\pi) (\bar X- \mu_X)^\top \beta_0 +o_p(n^{-1/2})\nonumber\\
	&= \underbrace{ \frac1n \sum_{i=1}^{n} I_i \{ \olim_{i1} - \theta_1- (X_i - \mu_X)^\top \beta_1  \}}_{M_1}-  \underbrace{\frac1n \sum_{i=1}^{n}(1-I_i)  \{\olim_{i0} - \theta_0 -  (X_i - \mu_X)^\top \beta_0\}}_{M_2} \nonumber \\
	&\qquad + \underbrace{(\bar X- E(\bar X\mid \mathcal{I}, \mathcal{S}))^\top (\pi\beta_1 - (1-\pi)\beta_0)}_{M_3} +  \underbrace{(E(\bar X\mid \mathcal{I}, \mathcal{S})-\mu_X)^\top (\pi\beta_1 - (1-\pi)\beta_0)}_{M_4}  \nonumber  \\
	&\qquad +o_p(n^{-1/2}) \nonumber\\
	&:= M_1 - M_2 + M_3 + M_4 +o_p(n^{-1/2}), \nonumber
\end{align}

By using the definition $ \beta_j =  \Sigma_X^{-1} {\mathrm cov} (X_i, \olim_{ij})$, we have 
\begin{align*}
	E\big[ X_i^\top \{ \olim_{ij} - \theta_j - (X_i - \mu_X)^\top \beta_j  \} \big] = {\mathrm cov} (X_i, \olim_{ij} ) -  {\mathrm cov} (X_i, \olim_{ij} )  = 0
\end{align*}  
Because $ Z_i $ is discrete and $ X_i $ contains all joint levels of $ Z_i $ as a sub-vector, according to the estimation equations from the least squares, we have that 
\begin{align*}
	E\left[ I(Z_i =z) \{ \olim_{ij}   - \theta_j 
	- (X_i - \mu_X)^\top \beta_j  \}\right] = 0, ~ \forall z\in \mathcal{Z},
\end{align*}
and thus, 
\begin{align}
	E\left[  \olim_{ij}  - \theta_j  - (X_i - \mu_X)^\top \beta_j  \mid Z_i \right] = 0,   ~ \text{a.s..}  \label{eq: conditional exp}
\end{align}

Next, we show that $ \sqrt{n} (M_1- M_2 +M_3  ) $ is asymptotically normal.  Consider the  random vector 
\begin{align}
	&\sqrt{n} \left( \begin{array}{c}
		E_n \left[ I_i (\olim_{i1}   - \theta_1 - (X_i - \mu_X )^\top \beta_1 )\right] \\
		E_n \left[ (1- I_i) (\olim_{i0}   - \theta_0 - (X_i - \mu_X )^\top \beta_0 )\right]  \\ 
		E_n \left[ (X_i  -  E(X_i \mid Z_i ))\right] 
	\end{array}
	\right), \label{eq: random vector}
\end{align}
where $ E_n [K_i] : = \frac{1}{n} \sum_{i=1}^{n} K_i$. Conditional on $\mathcal{I}, \mathcal{S}$, every component in \eqref{eq: random vector} is an average of independent terms. Similar to the proof of Theorem 2 in \cite{ye2021better}, the Lindeberg's Central Limit Theorem justifies that
\eqref{eq: random vector}  is asymptotically normal with mean 0 conditional on $\mathcal{I}, \mathcal{S}$,  as $ n\rightarrow \infty $. This implies that  $ \sqrt{n} (M_1- M_2 +M_3  ) $  is asymptotically normal with mean 0 conditional on $\mathcal{I}, \mathcal{S}$. Then, we calculate its variance. Note that 
\begin{align*}
	\var \left( \sqrt{n} (M_1- M_2)\mid \mathcal{I}, \mathcal{S}\right) 
	&= \frac1n \sum_{i=1}^{n}  I_i \var (\olim_{i1} - \theta_1 - (X_i- \mu_X)^\top \beta_1 \mid Z_i )  \\
	& \quad  +\frac1n \sum_{i=1}^{n}  (1-I_i) \var (\olim_{i0} - \theta_0 - (X_i- \mu_X)^\top \beta_0 \mid Z_i )  \\
	&=  \frac1n \sum_z \sum_{i: Z_i =z}  I_i \var (\olim_{i1} - \theta_1 - (X_i- \mu_X)^\top \beta_1 \mid Z_i =z )  \\
	& \quad + (1-I_i) \var (\olim_{i0} - \theta_0 - (X_i- \mu_X)^\top \beta_0 \mid Z_i =z )  \\
	&=  \sum_z \frac{n_1(z)}{n} \var (\olim_{i1} - \theta_1 - (X_i- \mu_X)^\top \beta_1 \mid Z_i =z ) \\
	& \quad  + \frac{n_0(z)}{n} \var (\olim_{i0} - \theta_0 - (X_i- \mu_X)^\top \beta_0 \mid Z_i =z )  \\
	&=  \sum_z \frac{n_1(z)}{n(z)}\frac{n(z)}{n} \var (\olim_{i1} - \theta_1 - (X_i- \mu_X)^\top \beta_1 \mid Z_i =z )  \\
	& \quad +\frac{n_0(z)}{n(z)}\frac{n(z)}{n} \var (\olim_{i0} - \theta_0 - (X_i- \mu_X)^\top \beta_0 \mid Z_i =z )  \\
	&= \pi  \sum_z  P(Z=z) \var (\olim_{i1} - \theta_1 - (X_i- \mu_X)^\top \beta_1 \mid Z_i =z )  \\
	&\quad +(1-\pi)  \sum_z  P(Z=z) \var (\olim_{i0} - \theta_0 - (X_i- \mu_X)^\top \beta_0 \mid Z_i =z ) + o_p(1)  \\
	&= \pi E\big\{ \var (\olim_{i1} - \theta_1 - (X_i- \mu_X)^\top \beta_1 \mid Z_i  ) \big\} \\
	& \quad + (1- \pi) E\big\{ \var (\olim_{i0} - \theta_0 - (X_i- \mu_X)^\top \beta_0 \mid Z_i  ) \big\}+ o_p(1)  \\
	&= \pi  \var (\olim_{i1} - \theta_1 - (X_i- \mu_X)^\top \beta_1 )\\
	& \quad +  (1- \pi)  \var (\olim_{i0} - \theta_0 - (X_i- \mu_X)^\top \beta_0 ) + o_p(1)  \\
	& =  \pi  \var (\olim_{i1}  - X_i^\top \beta_1 )+  (1- \pi)  \var (\olim_{i0}  - X_i^\top \beta_0 ) + o_p(1), 	
\end{align*}
$$ 	\var\left(\sqrt{n} \bar X \mid \mathcal{I}, \mathcal{S}\right)= \frac{1}{n} \sum_{i=1}^{n} \var (X_i \mid Z_i ) = E\{\var (X_i \mid Z_i )\} + o_p(1), $$
and
\begin{align*}
	n \cov(M_1 ,  \bar X \mid \mathcal{I}, \mathcal{S}) &= n \cov\left(\frac1n \sum_{i=1}^{n} I_i \{ \olim_{i1} - \theta_1- (X_i - \mu_X)^\top \beta_1  \},  \frac1n \sum_{i=1}^n  X_i \mid \mathcal{I}, \mathcal{S}\right)  \\
	&= \frac1n \sum_{i=1}^{n}I_i \cov\left(  \olim_{i1} - \theta_1- (X_i - \mu_X)^\top \beta_1  ,   X_i \mid Z_i\right)  \\
	&= \sum_z  \frac{n_1(z)}{n(z)} \frac{n(z)}{n}  \cov\left(  \olim_{i1} - \theta_1- (X_i - \mu_X)^\top \beta_1  ,   X_i \mid Z_i=z\right)  \\
	&=  \pi \sum_z  P(Z=z)\cov\left(  \olim_{i1} - \theta_1- (X_i - \mu_X)^\top \beta_1  ,   X_i \mid Z_i=z\right)  +o_p(1)\\
	&=  \pi E\left\{ \cov\left(  \olim_{i1} - \theta_1- (X_i - \mu_X)^\top \beta_1  ,   X_i \mid Z_i\right)  \right\} +o_p(1) \\
	&= o_p(1),
\end{align*}
where the last equality holds because $ E(\olim_{i1} - X_i^\top \beta_1 \mid Z_i ) = \theta_1- \mu_X^\top \beta_1 $ and, thus, $ \cov \{E(\olim_{i1} - X_i^\top \beta_1 \mid Z_i), E(X_i\mid Z_i) \} =0$ and $ E\big\{ \cov(  \olim_{i1} - \theta_1- (X_i - \mu_X)^\top \beta_1  ,   X_i \mid Z_i)  \big\} = \cov(  \olim_{i1} - \theta_1- (X_i - \mu_X)^\top \beta_1  ,   X_i ) =0   $ according to the definition of $ \beta_1 $. Similarly, we can show that $ n \cov(M_2 ,  \bar X \mid \mathcal{I}, \mathcal{S})= o_p(1)$. 

Combining the above derivations and from the Slutsky's theorem, we have shown that 
\begin{align*}
	\sqrt{n} (M_1- M_2+ M_3) \mid  \mathcal{I}, \mathcal{S} & \xrightarrow{d}  
	N \left( 0,   \pi  \var (\olim_{i1}  - X_i^\top \beta_1 )+  (1- \pi)  \var (\olim_{i0}  - X_i^\top \beta_0 )\right. \\
	& \quad \left.  + (\pi\beta_1 - (1-\pi)\beta_0)^\top E\{\var (X_i \mid Z_i )\} (\pi\beta_1 - (1-\pi)\beta_0) \right).
\end{align*}
From the bounded convergence theorem, this result also holds unconditionally, i.e.,  
\begin{align*}
	\sqrt{n} (M_1- M_2+ M_3) & \xrightarrow{d}  
	N \left( 0,   \pi  \var (\olim_{i1}  - X_i^\top \beta_1 )+  (1- \pi)  \var (\olim_{i0}  - X_i^\top \beta_0 ) \right. \\
	& \left. \quad + (\pi\beta_1 - (1-\pi)\beta_0)^\top E\{\var (X_i \mid Z_i )\} (\pi\beta_1 - (1-\pi)\beta_0) \right).
\end{align*}
Moreover, since $ M_4 $ is an average of i.i.d. terms, by the central limit theorem, 
\begin{align*}
	\sqrt{n} (E(\bar X \mid \mathcal{I}, \mathcal{S}) - \mu_X) = n^{-1/2} \sum_{i=1}^{n} \{ E(X_i\mid Z_i) - \mu_X\}\xrightarrow{d} N(0, \var (E(X_i\mid Z_i)) ), 
\end{align*}
and 
\begin{align*}
	\sqrt{n} M_4 \xrightarrow{d} N\big(0, (\pi\beta_1 - (1-\pi)\beta_0)^\top \var\{E (X_i \mid Z_i )\} (\pi\beta_1 - (1-\pi)\beta_0) \big)
\end{align*}
Next, we show that $ (\sqrt{n}(M_1- M_2+M_3), \sqrt{n} M_4) \xrightarrow{d}  (\xi_1, \xi_2)$, where $ (\xi_1, \xi_2) $ are mutually independent. This can be seen from 
\begin{align*}
	&P(\sqrt{n} (M_1- M_2+M_3) \leq t_1, \sqrt{n} M_4 \leq t_2) \\
	&= E\{I(\sqrt{n} (M_1- M_2+M_3) \leq t_1) I(\sqrt{n} M_4 \leq t_2)\} \\
	&= E\{P(\sqrt{n} (M_1- M_2+M_3) \leq t_1\mid \mathcal{I}, \mathcal{S}) I(\sqrt{n} M_4 \leq t_2)\} \\
	&= E\big[\{P(\sqrt{n} (M_1- M_2+M_3) \leq t_1\mid \mathcal{I}, \mathcal{S})  - P(\xi_1\leq t_1)\}I(\sqrt{n} M_4 \leq t_2)\big]  \\
	& \quad + P(\xi_1\leq t_1) P(\sqrt{n} M_4\leq t_2)\\
	&\rightarrow P(\xi_1\leq t_1) P(\xi_2\leq t_2),
\end{align*}
where the last step follows from the bounded convergence theorem. Finally, using the definitions of $\beta_0, \beta_1$,  it is easy to show that 
\begin{align*}
	&\pi  \var (\olim_{i1}  - X_i^\top \beta_1 )+  (1- \pi)  \var (\olim_{i0}  - X_i^\top \beta_0 ) + (\pi\beta_1 - (1-\pi)\beta_0)^\top \Sigma_X (\pi\beta_1 - (1-\pi)\beta_0) \\
	&=\pi \var (O_{i1}) + (1-\pi) \var(O_{i0})  - \pi(1-\pi) (\beta_1+\beta_0)^\top \Sigma_X (\beta_1+\beta_0), 
\end{align*}
concluding the proof that 
\begin{align*}
	& \sqrt{n}  \left\{ \hat  U_{\mathrm {CL}} -  \left(\frac{n_1}{n} \theta_1 -\frac{n_0}{n} \theta_0 \right) \right\}  \xrightarrow{d} N(0,\sigma_{\mathrm {CL}}^2 ). 
\end{align*}

\noindent
(b)  Since 
\begin{align*}
	\theta_1 =  E(\olim_{i1} ) & = \int_0^\tau \{ 1- \mu(t)\} \left[  E\{ dN_{i1}(t)\} - E\{Y_{i1} (t)\} p(t)dt \right] \\
	&= \int_0^\tau \{ 1- \mu(t)\} \left[  E\{ Y_{i1}(t)\lambda_1(t, V_i)\} dt - E\{Y_{i1} (t)\} p(t)dt \right] ,
\end{align*}
where $V_i=\emptyset$ under (C), and $V_i $ is the $V_i$ in (CR) under (CR)-(TR). Thus, the fact that	$\theta_1 = 0$ under $H_0$ follows from Lemma \ref{lemma: S3}(c). 
Similarly, $\theta_0= 0$ under $H_0$. 

Next, $\hat\sigma_{\mathrm {CL}}^2 \xrightarrow{p}  \sigma_{\mathrm {CL}}^2$ under $H_0$ is from $\hat\sigma_{\mathrm {L}}^2 \xrightarrow{p}  \sigma_{\mathrm {L}}^2$ as shown in the Proof of (17) in \cite{Ye:2020survival}, and 
$\hat\beta_j= \beta_j+ o_p(1), j=0,1$ and $\hat\Sigma_X = \Sigma_X + o_p(1)$ from Lemma \ref{lemma: beta}. The result that ${\cal T}_{\mathrm {CL}} \xrightarrow{d} N(0,1)$  follows from Slutsky's theorem.  

\noindent
(c) Under the local alternative, from Lemma \ref{lemma: S1}, we have that $\hat\sigma_{\mathrm {CL}}^2 = \sigma_{\mathrm {CL}}^2 +o_p(1) $, and thus 
%
\begin{align*}
	{\cal T}_{\mathrm {CL}} - \frac{\pi c_1 - (1-\pi) c_0}{\sigma_{\mathrm {CL}}} & = \frac{\sqrt{n} \hat U_{\mathrm {CL}}}{\hat \sigma_{\mathrm {CL}}} - \frac{\pi c_1 - (1-\pi) c_0}{\sigma_{\mathrm {CL}}}   \\
	&= \frac{\sqrt{n} \left( \hat U_{\mathrm {CL}} - \frac{n_1\theta_1 - n_0 \theta_0}{n} \right)}{\hat \sigma_{\mathrm {CL}}} + \frac{n_1c_1 - n_0 c_0 }{n\hat \sigma_{\mathrm {CL}}}  - \frac{\pi c_1 - (1-\pi) c_0}{\sigma_{\mathrm {CL}}}  \\
	&=  \frac{\sqrt{n} \left( \hat U_{\mathrm {CL}} - \frac{n_1\theta_1 - n_0 \theta_0}{n} \right)}{\sigma_{\mathrm {CL}}}  +o_p(1)\\
	&\xrightarrow{d} N(0,1).
\end{align*}

\noindent
{\sc Proof of Theorem S1}. 

\noindent
Similar to the Taylor expansion in the proof of Theorem 1(a), 
\begin{align*}
	&\hat U_{\mathrm {CL}} (b_0, b_1)  - \frac{n_1\theta_1 - n_0\theta_0}{n}  \\
	&= \frac{1}{n} \sum_{i=1}^{n} \left[ I_i  \{ \olim_{i1} - \theta_1 - (X_i- \mu_X )^\top b_1 \} -  (1-I_i) \{ \olim_{i0}  - \theta_0 - (X_i- \mu_X)^\top b_0\}   \right]\\
	&\qquad  + \frac{n_1}{n} (\bar X - \mu_X)^\top b_1   -\frac{n_0}{n} (\bar X - \mu_X)^\top b_0 + o_p(n^{-1/2})   \\
	&=  \underbrace{\frac{1}{n} \sum_{i=1}^{n} I_i \left[  \{ \olim_{i1} - \theta_1 - (X_i- \mu_X )^\top b_1  \}  - E\{ \olim_{i1} - \theta_1 - (X_i- \mu_X )^\top b_1  \mid Z_i\}    \right]}_{M_1} \\
	&\qquad -  \underbrace{\frac{1}{n} \sum_{i=1}^{n} (1-I_i) \left[  \{ \olim_{i0}  - \theta_0 - (X_i- \mu_X)^\top b_0\}   - E\{ \olim_{i0} - \theta_0 - (X_i- \mu_X )^\top b_0  \mid Z_i\}  \right]}_{M_2}\\
	&\qquad +\underbrace{  \bigg (\bar X - \frac{1}{n} \sum_{i=1}^{n} E(X_i \mid Z_i )  \bigg)^\top \left( \pi  b_1   - (1-\pi) b_0  \right) }_{M_3}  \\
	&\qquad +   \underbrace{\frac{1}{n} \sum_{i=1}^{n} (I_i - \pi)  \left[ E\{ \olim_{i1} - \theta_1 - (X_i- \mu_X )^\top b_1  \mid Z_i\}   + E\{ \olim_{i0} - \theta_0- (X_i- \mu_X )^\top b_0  \mid Z_i\}  \right]}_{M_4} \\
\end{align*}
\begin{align*}
	&\qquad  +  \underbrace{\frac{1}{n} \sum_{i=1}^{n} \left[   \pi E\{ \olim_{i1} - \theta_1 - (X_i- \mu_X )^\top b_1  \mid Z_i\}   -    (1-\pi) E\{ \olim_{i0} - \theta_0- (X_i- \mu_X )^\top b_0  \mid Z_i\} \right]}_{M_5} \\
	&\qquad  + \underbrace{  \bigg ( \frac{1}{n} \sum_{i=1}^{n} E(X_i \mid Z_i ) - \mu_X \bigg)^\top \left( \pi  b_1   - (1-\pi) b_0  \right) }_{M_6}+ o_p(n^{-1/2})    .
\end{align*}
In what follows, we will analyze these terms separately. Note that 
\begin{align*}
	n \var (M_1- M_2\mid \mathcal{I}, \mathcal{S}) &= \pi  E\left\{ \var ( \olim_{i1} - X_i^\top b_1  \mid Z_i)  \right\}  \\
	& \quad + (1- \pi ) E\left\{ \var ( \olim_{i0} - X_i^\top b_0  \mid Z_i)  \right\}  + o_p(1)  \\
	n \var (M_3  \mid   \mathcal{I}, \mathcal{S}) &=   \left( \pi  b_1   - (1-\pi) b_0  \right)^\top E \{ \var(X_i \mid Z_i )\} \left( \pi  b_1   - (1-\pi) b_0  \right) \\
	n \cov  (M_1- M_2, M_3 \mid   \mathcal{I}, \mathcal{S}) &= \{\pi (\beta_1 - b_1) - (1- \pi) (\beta_0 - b_0)\} ^\top E\{ \var (X_i \mid Z_i )\}  (\pi b_1 - (1-\pi)b_0) .
\end{align*}
Similar to the proof of Theorem 3 in \cite{ye2021better}, the Lindeberg's Central Limit Theorem and Slutsky theorem justify that $
\sqrt{n} (M_1-M_2+M_3)$ is asymptotically normal with mean 0 conditional on $ \mathcal{I}, \mathcal{S} $. Namely,   
\begin{align*}
	\sqrt{n} (M_1-M_2+M_3) \mid   \mathcal{I}, \mathcal{S} & \xrightarrow{d}  N\bigg( 0,   \pi  E\big\{ \var ( \olim_{i1} - X_i^\top b_1  \mid Z_i)  \big\}  \\
	& \quad + (1- \pi ) E\big\{ \var ( \olim_{i0} - X_i^\top b_0  \mid Z_i)  \big\} +  \{ 2\pi \beta_1- 2(1-\pi)\beta_0    \\
	&\quad 	- \pi b_1+ (1-\pi) b_0    \}^\top E\{ \var (X_i \mid Z_i )\} (\pi b_1 - (1-\pi)b_0)   \bigg).
\end{align*}
Moreover, from (C4),  $\sqrt{n}M_4$ is asymptotically normal conditional on $\mathcal{S}$, i.e., 
\begin{align*}
	\sqrt{n} M_4 \mid \mathcal{S} \xrightarrow{d} N \left (0, \nu  \var \big\{E ( \olim_{i1} - X_i^\top b_1  \mid Z_i)  + E( \olim_{i0} - X_i^\top b_0  \mid Z_i)   \big\} \right). 
\end{align*}
Because $M_5, M_6$ only involve sums of identically and independently distributed terms,  $E(M_5+M_6) = 0$, and
\begin{align*}
	n \var (M_5 )&=   \var \left\{ \pi E ( \olim_{i1} - X_i^\top b_1  \mid Z_i)  - (1-\pi) E( \olim_{i0} - X_i^\top b_0  \mid Z_i)   \right\}  \\
	n \var (M_6)&=  \left( \pi  b_1   - (1-\pi) b_0  \right)^\top \var \{ E(X_i \mid Z_i )\} \left( \pi  b_1   - (1-\pi) b_0  \right) \\
	n \cov (M_5, M_6)&=  \{\pi (\beta_1 - b_1) - (1- \pi) (\beta_0 - b_0)\} ^\top \var\{ E (X_i \mid Z_i )\}  (\pi b_1 - (1-\pi)b_0) ,
\end{align*}
we therefore have 
\begin{align*}
	\sqrt{n} (M_5+M_6) \xrightarrow{d} &	 N \left (0,  \var \left\{ \pi E ( \olim_{i1} - X_i^\top b_1  \mid Z_i)  - (1-\pi) E( \olim_{i0} - X_i^\top b_0  \mid Z_i)   \right\} \right. \\
	& \left.	  \{ 2\pi \beta_1- 2(1-\pi)\beta_0 - \pi b_1 + (1-\pi) b_0    \}^\top \var\{ E (X_i \mid Z_i )\} (\pi b_1 - (1-\pi)b_0)   \right). 
\end{align*}
Combining all the above derivations and similarly to the proof of Theorem 1, we can show that 
$ (\sqrt{n}(M_1- M_2+M_3), \sqrt{n} M_4, \sqrt{n} (M_5+M_6)) \xrightarrow{d}  (\xi_1, \xi_2, \xi_3)$, where $ (\xi_1, \xi_2, \xi_3) $ are mutually independent. Therefore,  $\sqrt{n} \left( \hat U_{\mathrm {CL}} (b_0, b_1)  - \frac{n_1\theta_1 - n_0\theta_0}{n}\right)$ is asymptotically normal with mean 0 and variance 
\begin{align*}
	\sigma_{\mathrm {L}}^2 (b_0, b_1) = & 	 \pi  E\left\{ \var ( \olim_{i1} - X_i^\top b_1  \mid Z_i)  \right\}  + (1- \pi ) E\left\{ \var ( \olim_{i0} - X_i^\top b_0  \mid Z_i)  \right\}  \\
	&\qquad +   \nu  \var \left\{E ( \olim_{i1} - X_i^\top b_1  \mid Z_i)  + E( \olim_{i0} - X_i^\top b_0  \mid Z_i)   \right\}    \\
	&\qquad +  \var \left\{ \pi E ( \olim_{i1} - X_i^\top b_1  \mid Z_i)  - (1-\pi) E( \olim_{i0} - X_i^\top b_0  \mid Z_i)   \right\}   \\
	&\qquad 	 +  \{ 2\pi \beta_1- 2(1-\pi)\beta_0 - \pi b_1 + (1-\pi) b_0    \}^\top \var (X_i  ) (\pi b_1 - (1-\pi)b_0).   
\end{align*}
Let $\sigma^2_{\mathrm {CL, SR}} (\beta_0, \beta_1) $ be the asymptotic variance under simple randomization. From the fact that $ \sigma^2_{\mathrm {CL}} (\beta_0, \beta_1) =\sigma^2_{\mathrm {CL, SR}} (\beta_0, \beta_1) $, we have 
\begin{align*}
	\sigma^2_{\mathrm {CL}} (b_0, b_1) - 	\sigma^2_{\mathrm {CL}} (\beta_0, \beta_1) &= 	\sigma^2_{\mathrm {CL}} (b_0, b_1)  - 	\sigma^2_{\mathrm CL, SR} (b_0, b_1)  + \sigma^2_{\mathrm {CL, SR}} (b_0, b_1)  - \sigma^2_{\mathrm {CL, SR}} (\beta_0, \beta_1). 
\end{align*}
Note first that 
\begin{align*}
	&	\sigma^2_{\mathrm {CL}} (b_0, b_1)  - 	\sigma^2_{\mathrm {CL, SR}} (b_0, b_1) \\
	&= - \{ \pi(1-\pi) - \nu \} \var \left\{ E(O_{i1 } - X_i^\top b_1 \mid Z_i ) + E(O_{i0 } - X_i^\top b_0 \mid Z_i ) \right\}\\
	&= - \{ \pi(1-\pi) - \nu \} (\beta_1 - b_1 + \beta_0 - b_0 )^\top \var\{ E(X_i \mid Z_i)\} (\beta_1 - b_1 + \beta_0 - b_0 ),
\end{align*}
where the third line is because $X_i$ includes all joint levels of $Z_i$ and thus $ E(O_{i1} - X_i^\top \beta_1\mid Z_i ) = \theta_1 - \mu_X^\top \beta_1$. 

To calculate $ \sigma^2_{\mathrm {CL, SR}} (b_0, b_1)  - \sigma^2_{\mathrm {CL, SR}} (\beta_0, \beta_1)$, we first note that 
\begin{align*}
	\sigma^2_{\mathrm {CL, SR}} (b_0, b_1) &= \var \left[ I_i \{ \olim_{i1} - \theta_1 - (X_i - \mu_X)^\top b_1 \} - (1- I_i)\{ \olim_{i0} - \theta_0 - (X_i - \mu_X)^\top b_0 \} \right] \\
	&\qquad + \{ 2\pi \beta_1 - 2(1- \pi) \beta_0 - \pi b_1 +(1-\pi) b_0 \}^\top \Sigma_X \{ \pi b_1- (1-\pi) b_0\} \\
	&=  \var \bigg[ I_i \{ \olim_{i1} - \theta_1 - (X_i - \mu_X)^\top \beta_1+ (X_i - \mu_X)^\top (\beta_1 - b_1) \}  \\
	&\qquad - (1- I_i)\{ \olim_{i0} - \theta_0 - (X_i - \mu_X)^\top \beta_0 +  (X_i - \mu_X)^\top(\beta_0 - b_0)  \}  \bigg] \\
	&\qquad + \{ 2\pi \beta_1 - 2(1- \pi) \beta_0 - \pi b_1 +(1-\pi) b_0 \}^\top \Sigma_X \{ \pi b_1- (1-\pi) b_0\} \\
	&= \var \left[ I_i \{ \olim_{i1} - \theta_1 - (X_i - \mu_X)^\top \beta_1 \} - (1- I_i)\{ \olim_{i0} - \theta_0 - (X_i - \mu_X)^\top \beta_0 \} \right] \\
	&\qquad + \var \left[ I_i(X_i - \mu_X)^\top (\beta_1 -b_1) - (1- I_i) (X_i - \mu_X)^\top (\beta_0 - b_0)  \right]\\
	&\qquad + \{ 2\pi \beta_1 - 2(1- \pi) \beta_0 - \pi b_1 +(1-\pi) b_0 \}^\top \Sigma_X \{ \pi b_1- (1-\pi) b_0\} .
\end{align*}
Then, we can calculate that 
\begin{align*}
	&\sigma^2_{\mathrm {CL, SR}} (b_0, b_1)  - \sigma^2_{\mathrm {CL, SR}} (\beta_0, \beta_1) \\
	&= \var \left\{I_i(X_i - \mu_X)^\top (\beta_1 -b_1) - (1- I_i) (X_i - \mu_X)^\top (\beta_0 - b_0)  \right\} \\
	&\qquad  + \{ 2\pi \beta_1 - 2(1- \pi) \beta_0 - \pi b_1 +(1-\pi) b_0 \}^\top \Sigma_X \{ \pi b_1- (1-\pi) b_0\}  \\
	& \qquad -  \{ \pi \beta_1 - (1- \pi) \beta_0 \}^\top \Sigma_X  \{ \pi \beta_1 - (1- \pi) \beta_0 \} \\
	&=\var \left\{I_i(X_i - \mu_X)^\top (\beta_1 -b_1)  \right\} +\var \left\{ (1- I_i) (X_i - \mu_X)^\top (\beta_0 - b_0)  \right\} \\
	&\qquad  + \{ 2\pi \beta_1 - 2(1- \pi) \beta_0 - \pi b_1 +(1-\pi) b_0 \}^\top \Sigma_X \{ \pi b_1- (1-\pi) b_0\}  \\
	& \qquad -  \{ \pi \beta_1 - (1- \pi) \beta_0 \}^\top \Sigma_X  \{ \pi \beta_1 - (1- \pi) \beta_0 \} \\
	&= \pi (\beta_1- b_1)^\top \Sigma_X (\beta_1 -b_1) + (1-\pi) (\beta_0 - b_0)^\top \Sigma_X (\beta_0 - b_0) \\
	&\qquad + \{ \pi\beta_1 - (1-\pi) \beta_0 \}^\top  \Sigma_X \{ \pi b_1 - (1-\pi) b_0 \}\\
	&\qquad +\{\pi (\beta_1 - b_1) - (1-\pi) (\beta_0 - b_0)\}^\top \Sigma_X \{ \pi b_1 - (1-\pi) b_0 \} \\
	&\qquad - \{ \pi \beta_1 - (1-\pi) \beta_0\}^\top \Sigma_X \{ \pi \beta_1 - (1-\pi \beta_0 )\} \\
	&= \pi (1- \pi) (\beta_1 - b_1 +\beta_0 - b_0)^\top \Sigma_X (\beta_1 - b_1 +\beta_0 - b_0). 
\end{align*}
Combining aforementioned results, we conclude that 
\begin{align*}
	&\sigma^2_{\mathrm {CL}} (b_0, b_1)  - \sigma^2_{\mathrm {CL}} (\beta_0, \beta_1)\\
	&= \pi(1-\pi) (\beta_1 - b_1 + \beta_0 - b_0)^\top E\{ \var(X_i \mid Z_i)\} (\beta_1 - b_1 + \beta_0 - b_0) \\
	&\qquad + \nu (\beta_1 - b_1 + \beta_0 - b_0)^\top \var\{ E(X_i\mid Z_i)\} (\beta_1 - b_1 + \beta_0 - b_0),
\end{align*}
which is greater or equal to zero because  $ E\{ \var(X_i \mid Z_i)\} $ and $ \var\{ E(X_i \mid Z_i)\} $ are positive definite. \\

\noindent 
{\sc  Proof of Theorem \ref{theo: hazard estimation Cox model}}. 

Since $\hat \theta_{\mathrm{CL}}$ solves $	\hat U_{\mathrm{CL}} (\vartheta ) = 0$, from the standard argument of M-estimation, we will show that under the Cox model $\lambda_1(t) = \lambda_0 (t) e^{\theta}$, 
\begin{align}
	\sqrt{n} \hat U_{\mathrm{CL}}(\theta)  & \xrightarrow{d}  N(0, \sigma_{\mathrm{CL}}^2 (\theta)) \label{eq: UCL} \\
	-\partial \hat   U_{\mathrm{CL}}(\vartheta)/\partial \vartheta \mid_{\vartheta = \bar\vartheta}  & = 	-\partial \hat   U_{\mathrm{L}}(\vartheta)/\partial \vartheta \mid_{\vartheta = \bar\vartheta}  \xrightarrow{p}  \sigma_{\mathrm{L}}^2 (\theta)  \label{eq: UCL deriv} 
\end{align}
where $\bar\vartheta$ lies  between $\hat \theta_{\mathrm{CL}}$   and $\theta$. Therefore, 
$$
\sqrt{n} (\widehat\theta_{\mathrm{ CL}}  - \theta ) \xrightarrow{d} N \left ( 0,\frac{ \sigma_{\mathrm{CL}}^2 (\theta) }{\sigma^4_{\mathrm {L}} (\theta )} \right) . 
$$

We first consider \eqref{eq: UCL}. Following the steps in the proof of Theorem 1, we can linearize $\hat U_{\mathrm{L}}(\theta_0)$ and obtain 
\begin{align*}
	\hat U_{\mathrm{L}} (\theta ) & = \frac1n \sum_{i=1}^n \{  I_i O_{i1} (\theta) - (1-I_i) O_{i0} (\theta) \} + n^{-1/2} o_p(1), \\
	O_{i1} (\theta) & = \int_0^\tau \left\{ 1- \frac{s^{(1)} (\theta, t)}{s^{(0)} (\theta, t)}\right\} \left\{ dN_{i1}(t) - Y_{i1}(t)  e^{\theta} \frac{E(dN_{i}(t))}{s^{(0)} (\theta, t)}\right\} \\
	O_{i0} (\theta) & = \int_0^\tau  \frac{s^{(1)} (\theta, t)}{s^{(0)} (\theta, t)} \left\{ dN_{i0}(t) - Y_{i0}(t) \frac{E(dN_{i}(t))}{s^{(0)} (\theta, t)}\right\} 	
\end{align*}
where $s^{(1)}(\theta , t) = e^{\theta} \mu(t) E\{  Y_i(t)\} $ and $s^{(0)} (\theta, t) = \{ e^{\theta}  \mu(t) +1- \mu(t)\} E\{  Y_i(t)\}$. In addition, similar to the  proof of Lemma \ref{lemma: beta}, we can show that $\hat \beta_j(\hat\theta_{\mathrm{L}}) \xrightarrow{p} \beta_j (\theta)$ for $j=0,1$. Thus, 
\begin{align*}
	\hat U_{\mathrm {CL}} (\theta ) &  =   n^{-1} \sum_{i=1}^n \left[   I_i \{ \olim_{i1} (\theta )  - (X_i - \bar X)^\top \beta_1 (\theta)\}- (1- I_i)\{ \olim_{i0}  (\theta) - (X_i - \bar X)^\top \beta_0 (\theta) \} \right]  \\
	&\quad +o_p(n^{-1/2}).
\end{align*}
Then as $E\{\olim_{i1} (\theta )   \}  =E\{\olim_{i0} (\theta )   \}= 0$, we have 
\begin{align*}
	& \sqrt{n}  \hat  U_{\mathrm {CL}}(\theta )  \xrightarrow{d} N(0,\sigma_{\mathrm {CL}}^2 (\theta) ), 
\end{align*}
where 
\begin{align*}
	\sigma_{\mathrm {CL}}^2 (\theta)& = \pi  \var (\olim_{i1} (\theta)  - X_i^\top \beta_1(\theta)   )+  (1- \pi)  \var (\olim_{i0}  (\theta)  - X_i^\top \beta_0 (\theta)  ) \\
	&\quad + (\pi\beta_1(\theta)  - (1-\pi)\beta_0(\theta)  )^\top \Sigma_X (\pi\beta_1(\theta)  - (1-\pi)\beta_0(\theta)  ) \\
	&=\pi \var (O_{i1}(\theta) ) + (1-\pi) \var(O_{i0}(\theta) )  - \pi(1-\pi) (\beta_1(\theta) +\beta_0(\theta) )^\top \Sigma_X (\beta_1(\theta) +\beta_0(\theta) ). 
\end{align*}

For \eqref{eq: UCL deriv}, note that 
\begin{align*}
	- \frac{\partial \hat U_{\mathrm{CL}} (\vartheta) }{\partial \vartheta} \mid_{\vartheta = \bar \vartheta}  & =  - \frac{\partial \hat U_{\mathrm{L}} (\vartheta) }{\partial \vartheta} \mid_{\vartheta = \bar \vartheta} = n^{-1} \sum_{i=1}^n \int_0^\tau  \left\{ \frac{S^{(1)} (\bar \vartheta, t) }{S^{(0)} (\bar\vartheta, t)}\right\}  -  \left\{ \frac{S^{(1)} (\bar \vartheta, t) }{S^{(0)} (\bar\vartheta, t)}\right\}^2 d N_i(t) \\
	& \xrightarrow{p} \int_0^\tau  \left\{ \frac{  s^{(1)} (\theta , t)  }{s^{(0)} (\theta, t)}-  \frac{s^{(1)} (\theta, t) ^2 }{s^{(0)} (\theta, t)^2}\right\}    s^{(0)} (\theta, t) dt .
\end{align*}
It remains to verify that 
\begin{align*}
	\pi \var (O_{i1}(\theta) ) + (1-\pi) \var(O_{i0}(\theta) )  =  \int_0^\tau  \left\{ \frac{  s^{(1)} (\theta , t)  }{s^{(0)} (\theta, t)}-  \frac{s^{(1)} (\theta, t) ^2 }{s^{(0)} (\theta, t)^2}\right\}    s^{(0)} (\theta, t) dt , 
\end{align*}
which is easy to show from $E(O_{i1} (\theta)) = 0$ and 
\begin{align*}
	\var (O_{i1}(\theta) ) & = 	E \{ O_{i1}^2 (\theta)\} =  \int_0^\tau \left\{ 1- \frac{s^{(1)} (\theta, t)}{s^{(0)} (\theta, t)}\right\}^2   E\{ dN_{i1}(t)\},  \\
	\var (O_{i0}(\theta) ) & = 	E \{ O_{i0}^2 (\theta)\} =  \int_0^\tau \left\{  \frac{s^{(1)} (\theta, t)}{s^{(0)} (\theta, t)}\right\}^2   E\{ dN_{i0}(t)\} .
\end{align*}

\noindent
{\sc Proof of Theorem 2}.

\noindent
Part (a) is from Theorem \ref{theo: b0 b1}(a) with $b_0=b_1=0$. 
For part (b), $\theta_1= \theta_0= 0$ is proved in Theorem 1(b), $\hat\sigma_{\mathrm {L}}^2 \xrightarrow{p} \sigma_{\mathrm {L}}^2 $  is proved in  \cite{Ye:2020survival}. 
For part (c), note that  $\hat\sigma_{\mathrm {L}}^2 =\sigma_{\mathrm {L}}^2  + o_p(1) $ under the local alternative  (Lemmas \ref{lemma: S0}-\ref{lemma: S1}). Hence,
\begin{align*}
	{\cal T}_{\mathrm {L}} - \frac{\pi c_1 - (1-\pi) c_0}{\sigma_{\mathrm {L}}} & = \frac{\sqrt{n} \hat U_{\mathrm {L}}}{\hat \sigma_{\mathrm {L}}} - \frac{\pi c_1 - (1-\pi) c_0}{\sigma_{\mathrm {L}}}   \\
	&= \frac{\sqrt{n} \left( \hat U_{\mathrm {L}} - \frac{n_1\theta_1 - n_0 \theta_0}{n} \right)}{\hat \sigma_{\mathrm {L}}} + \frac{n_1c_1 - n_0 c_0 }{n\hat \sigma_{\mathrm {L}}}  - \frac{\pi c_1 - (1-\pi) c_0}{\sigma_{\mathrm {L}}}  \\
	&=  \frac{\sqrt{n} \left( \hat U_{\mathrm {L}} - \frac{n_1\theta_1 - n_0 \theta_0}{n} \right)}{\sigma_{\mathrm {L}}}  +o_p(1)\\
	&\xrightarrow{d} N \left(0,\frac{\sigma_{\mathrm {L}}^2(\nu) }{\sigma_{\mathrm {L}}^2 }\right).
\end{align*}

\noindent
{\sc Proof of Theorem 3}. 

\noindent
(a) From linearizing $	\hat	U_{\mathrm SL} $ \citep{Ye:2020survival}, we have that 
\begin{align}
	& \hat	U_{\mathrm SL} -   \sum_z \left(\frac{n_{z1} }{n} \theta_{z1} -\frac{n_{z0} }{n} \theta_{z0} \right) \nonumber \\
	&  =\frac1n \sum_z \sum_{i: Z_i= z}   I_i  \left( \olim_{zi1} - \theta_{z1} \right)  -  (1-I_i)   \left( \olim_{zi0}  - \theta_{z0}\right)+o_p(n^{-1/2}).  \nonumber
\end{align}
Following similar  steps as in the proof of Theorem 1, we have that 
\begin{align*}
	\sqrt{n} \left\{  \hat	U_{\mathrm SL} -    \sum_z \left(\frac{n_{z1} }{n} \theta_{z1} -\frac{n_{z0} }{n} \theta_{z0} \right)
	\right\} \xrightarrow{d} N\left(0,    \sigma_{\mathrm SL}^2 \right). 
\end{align*}

For the calibrated stratified log-rank test $ \hat U_{{\mathrm CSL}} $, from the linearization of $\hat U_{\mathrm SL}$,
$n^{-1} \sum_z \sum_{i:Z_i=z} I_i (X_i - \bar X_z) = O_p(n^{-1/2})$, and  $ \bar X_z - E(X_i \mid Z_i = z) = O_p(n^{-1/2}) $,   we have 
\begin{align}
	&\hat U_{\mathrm CSL} -     \sum_z \left(\frac{n_{z1} }{n} \theta_{z1} -\frac{n_{z0} }{n} \theta_{z0} \right)  \nonumber \\
	&  =\sum_z \frac1n \sum_{i=1}^n    I_i  {\cal I}(Z_i=z) \left( \olim_{zi1} - \theta_{z1} - (X_i - E(X_i\mid Z_i=z) )^\top\gamma_1 \right)  \nonumber \\
	&\qquad -\sum_z \frac1n \sum_{i=1}^n  (1-I_i)  {\cal I}(Z_i=z) \left( \olim_{zi0}  - \theta_{z0} - (X_i -E(X_i\mid Z_i=z) )^\top \gamma_0 \right) \nonumber\\
	&\qquad  + \sum_z \P (Z=z) (\pi\gamma_1 - (1-\pi) \gamma_0)^\top (\bar X_z - E(X_i\mid Z_i=z) )+o_p(n^{-1/2}).  \nonumber \\
	&:= B_1-B_2+B_3 +o_p(n^{-1/2}). 
\end{align}
Next, we show that $ \sqrt{n} (B_1-B_2+B_3 )  $ is asymptotically normal. Let $\mu_{Xz} = E(X_i\mid Z_i=z)$. Consider the random vector 
\begin{align}
	&\sqrt{n} \left( \begin{array}{c}
		\left( E_n \big[ I_i {\cal I}(Z_i =z)(\olim_{zi1}   - \theta_{z1} - (X_i - \mu_{Xz} )^\top \gamma_1 )\big], z\in \mathcal{Z}  \right)^\top \\
		\left( E_n \big[ (1-I_i) {\cal I}(Z_i =z)(\olim_{zi0}   - \theta_{z0} - (X_i - \mu_{Xz} )^\top \gamma_0 )\big], z\in \mathcal{Z}  \right)^\top\\
		\left(	E_n \big[{\cal I}(Z_i=z) (X_i  - \mu_{Xz}  )\big],  z\in \mathcal{Z} \right)^\top
	\end{array}
	\right). \label{eq: SL random vector}
\end{align}
Conditional on $ \mathcal{I}, \mathcal{S} $, every component in \eqref{eq: SL random vector} is an average of independent terms. From Lindeberg's Central Limit Theorem, as $ n\rightarrow\infty $, \eqref{eq: SL random vector} is asymptotically normal with mean 0 conditional on  $ \mathcal{I}, \mathcal{S}$. This implies that $   \sqrt{n} (B_1-B_2+B_3 )  $ is asymptotically normal with mean 0 conditional on $ \mathcal{I}, \mathcal{S}$. Then, we calculate its variance. Note that 
\begin{align*}
	&\var (\sqrt{n} (B_1- B_2) \mid \mathcal{I}, \mathcal{S}  ) \\
	&= \sum_z \frac1n \sum_{i=1}^n I_i {\cal I} (Z_i = z) \var (\olim_{zi1}  - (X_i - \mu_{Xz} )^\top \gamma_1\mid Z_i=z ) \\
	&\quad + \sum_z \frac1n \sum_{i=1}^n (1-I_i) {\cal I}  (Z_i = z) \var (\olim_{zi0}  - (X_i - \mu_{Xz} )^\top \gamma_0 \mid Z_i=z )  \\
	&= \pi \sum_z P(Z_i=z) \var (\olim_{zi1}  - (X_i - \mu_{Xz} )^\top \gamma_1\mid Z_i=z )   \\
	&\quad +(1- \pi) \sum_z P(Z_i=z)\var (\olim_{zi0}  - (X_i - \mu_{Xz} )^\top \gamma_0 \mid Z_i=z ) +o_p(1) \\
	&= \pi \sum_z P(Z_i= z) \big\{ \var (\olim_{zi1 }\mid Z_i=z )  + \gamma_1^\top \var(X_i\mid Z_i=z) \gamma_1
	- 2 \gamma_{1}^\top  \cov(X_i, O_{zi1}\mid Z_i= z)\big\} \\
	&\quad + (1-\pi)  \sum_z P(Z_i= z) \big\{ \var (\olim_{zi0 }\mid Z_i=z )  + \gamma_0^\top \var(X_i\mid Z_i=z) \gamma_0 \\
	& \quad 
	- 2 \gamma_{0}^\top  \cov(X_i, O_{zi0}\mid Z_i= z)\big\} +o_p(1)\\
	&= \sigma_{\mathrm SL}^2 - \pi \gamma_1^\top E\{ \var (X_i \mid Z_i )\} \gamma_1 - (1-\pi) \gamma_0^\top E\{ \var (X_i \mid Z_i )\} \gamma_0    +o_p(1), 
\end{align*} 
where the last line is because 
\begin{align*}
	\gamma_j  = \bigg\{  \sum_z P(Z_i = z) \var (X_i \mid Z_i=z)\bigg\}^{-1} \bigg\{\sum_z P(Z_i = z) \cov (X_i, O_{zij} \mid Z_i=z)  \bigg\}
\end{align*}
for $j=0,1$, from  Lemma \ref{lemma: beta}. 
Moreover, 
\begin{align*}
	&\var (\sqrt{n} B_3\mid \mathcal{I}, \mathcal{S}  )\\
	&= (\pi\gamma_1- (1-\pi) \gamma_0)^\top E\{ \var (X_i\mid Z_i) \} (\pi\gamma_1- (1-\pi) \gamma_0) +o_p(1). 
\end{align*}
Next, we can calculate the covariance between $ \sqrt{n} B_1 $ and $ \sqrt{n} B_3 $ when conditional on $ \mathcal{I}, \mathcal{S}$ as
\begin{align*}
	& \cov (\sqrt{n} B_3, \sqrt{n} B_1\mid \mathcal{I}, \mathcal{S}) \\
	&= n\cov \left(  \sum_z \Pr(Z= z) (\pi\gamma_1- (1-\pi) \gamma_0)^\top (\bar X_z- \mu_{Xz}), \right. \\
	& \quad \left. \sum_z \frac{1}{n} \sum_{i=1}^n I_i {\cal I}(Z_i=z) (\olim_{zi1} - (X_i - \mu_{Xz} )^\top \gamma_1  ) \mid\mathcal{I}, \mathcal{S} \right) \\
	&= \sum_z \sum_{i=1}^n I_i {\cal I}(Z_i=z) \Pr(Z= z) (\pi\gamma_1- (1-\pi) \gamma_0)^\top \cov (\bar X_z - \mu_{Xz} , \olim_{zi1} -  (X_i - \mu_{Xz} )^\top \gamma_1 \mid \mathcal{I}, \mathcal{S} ) \\
	&= \sum_z  \sum_{i=1}^n I_i {\cal I}(Z_i=z) \Pr(Z= z) (\pi\gamma_1- (1-\pi) \gamma_0)^\top\frac{1}{n_z}\cov (X_i- \mu_{Xz} , \olim_{zi1} -  (X_i - \mu_{Xz} )^\top \gamma_1 \mid \mathcal{I}, \mathcal{S} ) \\
	&= \sum_z \pi \Pr(Z_i=z)  (\pi\gamma_1 - (1-\pi) \gamma_0 )^\top \left\{  \cov( X_i ,\olim_{zi1}\mid Z_i )- \var (X_i\mid Z_i) \gamma_1 \right\} +o_p(1) \\
	&= o_p(1)
\end{align*}
Similarly, $  \cov (\sqrt{n} B_3, \sqrt{n} B_2\mid \mathcal{I}, \mathcal{S}) = o_p(1) . $ Hence, $ \cov (\sqrt{n} B_3, \sqrt{n} (B_1- B_2)\mid \mathcal{I}, \mathcal{S}) =  o_p(1) $. 
Combining all the above derivations, we have that 
\begin{align*} 
	&\var \{\sqrt{n} (B_1-B_2+B_3 ) \mid  \mathcal{I}, \mathcal{S} \}  \\
	&= \sigma_{\mathrm SL}^2 - \pi(1-\pi) (\gamma_1 + \gamma_0)^\top E\{ \var (X_i\mid Z_i) \} (\gamma_1 + \gamma_0) +o_p(1). 
\end{align*}
The  asymptotic distribution then follows from the  Slutsky's theorem and bounded convergence theorem. 

\noindent
(b) It is straightforward to show the assumed conditions imply  $\theta_{z1} = \theta_{z0}=0$ for any $z$ from applying Lemma \ref{lemma: S3} separately for every  stratum $z$.

Next, under $H_0$, $\hat\sigma_{\mathrm SL}^2 \xrightarrow{p}  \sigma_{\mathrm SL}^2$ is proved in \cite{Ye:2020survival}, and  $\hat\sigma_{\mathrm CSL}^2 \xrightarrow{p}  \sigma_{\mathrm CSL}^2$ is from  
$\hat\gamma_j= \gamma_j+ o_p(1), j=0,1$ from Lemma \ref{lemma: beta},  $\hat\Sigma_{X\mid z} = \var (X_i \mid Z_i=z) + o_p(1)$, and $n_z/n= P(Z_i =z)+ o_p(1)$. 
The result that ${\cal T}_{\mathrm {CL}} \xrightarrow{d} N(0,1)$ and validity of ${\cal T}_{\mathrm {CL}}$ follows from Slutsky theorem.

\noindent
(c) Under the local alternative, applying Lemmas \ref{lemma: S0}-\ref{lemma: S1} within each stratum $Z_i=z$ gives $\hat\sigma_{\mathrm SL}^2 = \sigma_{\mathrm SL}^2 +o_p(1) $. In addition, from Lemma \ref{lemma: beta} and $\hat \Sigma_{X|z} = \var(X_i\mid Z_i=z) +o_p(1)$, we have  $\hat\sigma_{\mathrm CSL}^2 = \sigma_{\mathrm CSL}^2 +o_p(1) $. Then, 
\begin{align*}
	&{\cal T}_{\mathrm SL} - \frac{   \sum_z \Pr (Z= z) \{ \pi c_{z1} - (1-\pi) c_{z0}\}}{\sigma_{\mathrm SL}} \\
	& = \frac{\sqrt{n} \hat U_{\mathrm SL}}{\hat \sigma_{\mathrm SL}} - \frac{   \sum_z \Pr (Z= z) \{ \pi c_{z1} - (1-\pi) c_{z0}\}}{\sigma_{\mathrm SL}} \\
	&= \frac{\sqrt{n} \left( \hat U_{\mathrm SL} - \sum_z \frac{n_{z1}\theta_{z1} - n_{z0} \theta_{z0}}{n} \right)}{\hat \sigma_{\mathrm SL}} + \sum_z  \frac{n_{z1}c_{z1} - n_{z0} c_{z0}}{n \hat\sigma_{\mathrm SL}} - \frac{   \sum_z \Pr (Z= z) \{ \pi c_{z1} - (1-\pi) c_{z0}\}}{\sigma_{\mathrm SL}}  \\
	&= \frac{\sqrt{n} \left( \hat U_{\mathrm SL} - \sum_z \frac{n_{z1}\theta_{z1} - n_{z0} \theta_{z0}}{n} \right)}{\hat \sigma_{\mathrm SL}}+o_p(1)\\
	&\xrightarrow{d} N(0,1), 
\end{align*} 
and 
\begin{align*}
	&{\cal T}_{\mathrm CSL} - \frac{   \sum_z \Pr (Z= z) \{ \pi c_{z1} - (1-\pi) c_{z0}\}}{\sigma_{\mathrm CSL}} \\
	& = \frac{\sqrt{n} \hat U_{\mathrm CSL}}{\hat \sigma_{\mathrm CSL}} - \frac{   \sum_z \Pr (Z= z) \{ \pi c_{z1} - (1-\pi) c_{z0}\}}{\sigma_{\mathrm CSL}} \\
	&= \frac{\sqrt{n} \left( \hat U_{\mathrm CSL} - \sum_z \frac{n_{z1}\theta_{z1} - n_{z0} \theta_{z0}}{n} \right)}{\hat \sigma_{\mathrm CSL}} + \sum_z  \frac{n_{z1}c_{z1} - n_{z0} c_{z0}}{n \hat\sigma_{\mathrm CSL}} - \frac{   \sum_z \Pr (Z= z) \{ \pi c_{z1} - (1-\pi) c_{z0}\}}{\sigma_{\mathrm CSL}}  \\
	&= \frac{\sqrt{n} \left( \hat U_{\mathrm CSL} - \sum_z \frac{n_{z1}\theta_{z1} - n_{z0} \theta_{z0}}{n} \right)}{\hat \sigma_{\mathrm CSL}}+o_p(1)\\
	&\xrightarrow{d} N(0,1).
\end{align*}

\end{document}